\documentclass[superscriptaddress,twocolumn,
 amsmath,amssymb, aps,prb]{revtex4-2}
\usepackage{amsmath,amsfonts,amssymb,amsthm}
\usepackage{microtype}
\usepackage{graphicx}
\usepackage{float}
\usepackage{dcolumn}
\usepackage{bm}
\usepackage[bottom]{footmisc}
\usepackage{pdfpages}
\usepackage[caption=false]{subfig}
\usepackage{enumitem}
\usepackage[normalem]{ulem}
\makeatletter

\usepackage[colorlinks=true,
    linkcolor=blue,
    citecolor=red,
    filecolor=magenta,      
    urlcolor=cyan,
    pdftitle={Approximation Lanczos},]{hyperref}

\theoremstyle{plain}
\newtheorem{assumption}{Assumption}

\AtBeginDocument{\let\LS@rot\@undefined}
\makeatother

\begin{document}

\title{Approximation theory for Green's functions via the Lanczos algorithm}

\author{Gabriele Pinna} 
\affiliation{Department of Physics, King’s College London, Strand WC2R 2LS, UK}

\author{Oliver Lunt} 
\affiliation{Department of Physics, King’s College London, Strand WC2R 2LS, UK}
\affiliation{Department of Physics, University of Oxford, Parks Road, OX1 3PU, UK}

\author{Curt von Keyserlingk} 
 \affiliation{Department of Physics, King’s College London, Strand WC2R 2LS, UK}

\date{\today}
\begin{abstract}
It is known that Green's functions can be expressed as continued fractions; the content at the $n$-th level of the fraction is encoded in a coefficient $b_n$, which can be recursively obtained using the Lanczos algorithm. We present a theory concerning errors in approximating Green's functions using continued fractions when only the first $N$ coefficients are known exactly. Our focus lies on the stitching approximation (also known as the recursion method), wherein  truncated continued fractions are completed with a sequence of coefficients for which exact solutions are available. We assume a now standard conjecture about the growth of the Lanczos coefficients in chaotic many-body systems, and that the stitching approximation converges to the correct answer. Given these assumptions, we show that the rate of convergence of the stitching approximation to a Green's function depends strongly on the decay of staggered subleading terms in the Lanczos cofficients. Typically, the decay of the error term ranges from $1/\mathrm{poly}(N)$ in the best case to $1/\mathrm{poly}(\log N)$ in the worst case, depending on the differentiability of the spectral function at the origin.
We present different variants of this error estimate for different asymptotic behaviours of the $b_n$, and we also conjecture a relationship between the asymptotic behavior of the $b_n$'s and the smoothness of the Green's function. Lastly, with the above assumptions, we prove a formula linking the spectral function's value at the origin to a product of continued fraction coefficients, which we then apply to estimate the diffusion constant in the mixed field Ising model.
\end{abstract}

\maketitle

\section{Introduction} 

Understanding thermalization in many-body systems is a central problem in quantum physics \cite{d2016quantum,PhysRevA.43.2046,PhysRevE.50.888}. In particular, the study of transport of conserved quantities is a fundamental question of interest. The microscopic dynamics are unitary, but non-unitary hydrodynamical laws emerge at larger scales \cite{forster2018hydrodynamic, bulchandani2021superdiffusion, liu2018lectures}. Probing the hydrodynamic regime is a challenging task even computationally. As usual, the difficulties arise from the dimensionality of the Hilbert space which scales exponentially with the size of the system.  Recent developments in tensor networks mitigate the problem; in particular, matrix product states/operators allows for more efficient information storage \cite{PhysRevLett.91.147902}. The standard algorithm for dynamics is time-evolving block decimation (TEBD) \cite{PhysRevLett.93.040502}. However the required memory (and run-time) for TEBD is believed to grow exponentially in time, thus limiting the usefulness of TEBD for real-time hydrodynamical simulations. 

To address this issue, new methods have been introduced, including dissipation-assisted operator evolution (DAOE)\cite{PhysRevB.105.075131, PhysRevB.109.205108}, density matrix truncation (DMT) \cite{PhysRevB.97.035127}, and local-information time evolution (LITE) \cite{PRXQuantum.5.020352, 10.21468/SciPostPhys.13.4.080}.
An alternative technique to study dynamics involves the Lanczos algorithm, also known as the recurrence method, which transforms the Liouvillian into a tridiagonal form by constructing an orthonormal operator basis \cite{viswanath1994recursion}. This transformation effectively reduces the problem to a more familiar hopping model on a semi-infinite chain. The hopping terms in this tridiagonal matrix, denoted $b_n$ and called Lanczos coefficients, can then be used to represent the Green's function as a continued fraction from which it is possible in principle to extract the diffusion constant of the model \cite{PhysRevX.9.041017, PhysRevB.110.104413}. 

However, a significant source of error arises from the limited number of coefficients that can be extracted numerically. The primary challenge is related to memory constraints: in the worst-case scenario, the memory requirements grow exponentially with the number of Lanczos steps. Furthermore, it is not a priori clear what is the best way to approximate this continued fraction given only a limited number of coefficients. This problem of terminating a continued fraction has been investigated in the past; for instance, in the context of  approximating densities of states  for many-band Hamiltonians \cite{Haydock_1985}. Recently, the challenge of effectively approximating continued fractions has resurfaced in relation to thermalizing quantum systems. In this context, the operator growth hypothesis (OGH) \cite{PhysRevX.9.041017} proposes that in local thermalising quantum systems, the $b_n$ have a universal linear/quasilinear asymptotic growth with $n$ depending on the dimension. This conjecture has been proved to hold for the quantum Ising spin model in dimensions greater than one, and for the chaotic Ising chain in one dimension \cite{Cao_2021}. The OGH can be used to estimate a Green's function by approximating the asymptotic growth of the exact $b$ coefficients based on the first $N$ Lanczos coefficients extracted numerically. One then replaces the infinitely remaining set of $b$'s with a known (`Meixner–Pollaczek') sequence of Lanczos coefficients $b_{N+k}\rightarrow b_{N+k,s},k>0$ which have the same asymptotic behaviour (see Eq.~\eqref{eq:stitch}). It turns out that one can compute the resulting fraction exactly with this known sequence, and the result gives estimate for the Green's function.

We focus on the following question: how quickly does the error in the computed Green's function  decay to zero as a function of $N$? The results of this investigation  shed light on the computational efficiency of the Lanczos method, and quite possibly the inherent complexity of many-body dynamics.  We begin by investigating the general problem of approximating Green's functions using the Lanczos algorithm and provide some initial loose bounds at finite frequency.  We then do a more sophisticated analysis, and with some assumptions show that the error in the estimated Green's function depends strongly on the nature of the subleading corrections to the original operator growth hypothesis. Subsequently, we validate our analytical theory numerically using cases with known solutions. Finally, we apply the theory to a physical problem: extracting the diffusion constant in the chaotic Ising model. 

\section{Approximation of Green's functions}
This section investigates the general problem of approximating  Green's functions in unitary quantum systems by truncating/completing the above-mentioned continued fraction representation. 
Consider Hamiltonians $H$ that act on a Hilbert space $ \mathcal{H}$ with operator dynamics induced by the Liouvillian $\mathcal{L} = [H,\; \cdot \;]$. 
Let $O$ be a normalised operator, and define the corresponding Green's function as
\begin{equation}\label{G:resolvent}
    G(z) = \langle O, (z-\mathcal{L})^{-1}O \rangle \hspace{5pt},
\end{equation}
where $\langle \cdot, \cdot \rangle$ is the Kubo or thermal inner product \cite{forster2018hydrodynamic}. We also define the spectral function as
\begin{equation}
    \rho(x) = \frac{1}{\pi}\lim_{\epsilon \to 0^{+}}\Im\left(\langle O, (x-\mathrm{i} \epsilon-\mathcal{L})^{-1}O \rangle\right) ,
\end{equation}
where $\Im(\cdot)$ is the imaginary part. This is the continuous part of some spectral measure which we denote $d\mu(x) = \rho(x) dx$.

We are interested in the Green's function in the thermodynamic limit, when the dimensionality of the Hilbert space tends to infinity. To approximate this limit, we employ the Lanczos algorithm that brings a self-adjoint operator into a tridiagonal form. This is done by constructing an orthonormal Krylov basis $\{O_n\}_{n=0}^\infty$, defined via a three-term recurrence relation:
\begin{equation}
    b_{n+1}O_{n+1} = \mathcal{L}O_n - b_n O_{n-1} \hspace{5pt},
\label{eqn: Jacobi}
\end{equation}
with initial conditions $O_0 = O$, $O_{-1}=0$, and the convention $b_0 = 1$. Usually, the initial operator $O$ is Hermitian, which implies that the diagonal part of $\mathcal{L}$ vanishes, i.e., $\langle O_n, \mathcal{L}O_n \rangle = 0$. As a result, the algorithm yields:
\begin{enumerate}
    \item a sequence of Lanczos coefficients, $\{b_n\}_{n=1}^\infty$, determined by imposing, at each step, the normalization condition $\|O_n\| = 1$,
    \item an orthonormal Krylov basis $\{O_n\}_{n=0}^\infty$.
\end{enumerate}

When the system size, $L$, is finite, the Liouvillian $\mathcal{L}$ acts as a finite tridiagonal matrix; when $L \to \infty$, it acts as a Jacobi operator on square-summable sequences $\ell^2(\mathbb{Z}_{\geq 0})$ \cite{koelink2001spectraltheoryspecialfunctions}. 

Notice that the Lanczos operators $O_n$, in Eq.~\eqref{eqn: Jacobi}, can be expressed as $O_n = p_n(\mathcal{L})O_0$, where $\{p_n(x)\}$, by the spectral theorem, are orthonormal polynomials with respect to the measure (or weight) $\mu$ \cite{simon2015operator}. These polynomials satisfy the same recurrence relation in Eq.~\eqref{eqn: Jacobi}:
\begin{equation}
    b_{n+1}p_{n+1}(x) = x p_n(x) - b_n p_{n-1}(x) \hspace{5pt},
\label{eqn: Three-term recurrence: polynomials}
\end{equation}
with initial conditions $p_{-1}(x) = 0$ and $p_0(x) = 1$.

In theory, once all the Lanczos coefficients $\{b_n\}_{n=1}^\infty$ are determined via the three-term recurrence relation, they can be used to represent the Green’s function $G(z)$ through the continued fraction expansion \cite{PhysRevX.9.041017}: 
\begin{equation} G(z) = \cfrac{1}{z - \cfrac{b_1^2}{z - \cfrac{b_2^2}{z - \ddots}}} \hspace{5pt}.
\end{equation} 
However, in practice, only a limited number of coefficients are computable, so our focus is on approximating this expression using only the first $N$ coefficients.

The most drastic approximation is truncating the continued fraction by setting $b_{N+k}=0, \forall k>0$. The resulting expression converges when $\Im(z) \neq 0$, but the convergence worsens as $\Im(z)\rightarrow 0$ \cite{sm}. The heuristic explanation is that truncating the continued fraction corresponds to dealing with a finite-dimensional hopping problem for which the return probability does not decay with time. In contrast, we expect our Green's function to decay at late times in a many-body system in the thermodynamic limit. 

Consequently, we focus on a different method, the ``stitching" approximation, which mitigates the convergence issue \cite{viswanath1994recursion}. It consists of replacing the exact coefficients $\{b_{n}\}_{n=N+1}^{\infty}$, with a ``stitched" sequence $\{b_{n,s}\}_{n=N+1}^{\infty}$ for which the Green function $G_s(z)$ is known exactly. 
This approximation, unlike truncation, preserves the infinite dimensionality of the underlying vector space. In order to implement this numerically, it is useful to define a so-called $n$-th level Green's function $G^{(n)}(z)$. These functions are defined recursively by
\begin{equation}
    G^{(n)}(z) = M_{n+1} \left(G^{(n+1)}(z)\right) \equiv \frac{1}{z-b_{n+1}^2G^{(n+1)}(z)}, 
    \label{eqn: nth level Green's function}
\end{equation}
where $M_{n+1}$ is a M\"{o}bius transformation, and  we define $G^{(0)}(z) = G(z)$. The n-th level Green's function can be interpreted as a Green's function of the projected evolution operator
\begin{equation}
G^{(n)}(z) = \langle O_n, (z-\mathcal{P}_{\geq n}\mathcal{L}\mathcal{P}_{\geq n})^{-1} O_n \rangle \hspace{5pt},
\end{equation}
where \( \mathcal{P}_{\geq n} \) projects into the subspace spanned by the basis vectors \( \{O_j\}_{j=n}^{\infty} \). The physical interpretation of these Green's functions is related to the self energy $\Sigma(z)$, defined by
\begin{equation}
G(z) = \frac{1}{z-\Sigma(z)} \hspace{5pt}.    
\end{equation}
This relationship connects the first-level Green's function to the self-energy as follows: \begin{equation} 
\Sigma(z) = b^2_{1} G^{(1)}(z) = \langle \mathcal{L}O, (z - \mathcal{P}{\geq 1} \mathcal{L} \mathcal{P}{\geq 1})^{-1} \mathcal{L}O \rangle \hspace{5pt}. 
\end{equation} 
We can interpret the $n$-th level Green's function hierarchically, as being proportional to the self-energy associated with the preceding Green's function, $G^{(n-1)}(z)$. This formulation highlights the connection between the Lanczos algorithm and the memory function formalism \cite{forster2018hydrodynamic}, originally introduced by Mori \cite{10.1143/PTP.34.399}.

In terms of these functions, we may write the Green's function exactly as 
\begin{equation}\label{eq:iterative}
G(z)=\Omega_N\left(G^{(N)}(z)\right)
\end{equation}
where $\Omega_{N} =  M_1 \circ \ldots \circ M_N$ denotes the composition of $N$ M\"{o}bius transforms using the exact sequence $\{b_n\}_{n=1}^\infty$. Approximating $G(z)$ using stitching at level $N$ is equivalent to replacing $G^{(N)}(z)$, in Eq.~\eqref{eq:iterative}, with an analytically known Green's function obtained using the stitched Lanczos sequence, which we denote $G^{(N)}_s(z)$.
The stitching approximation is thus
\begin{equation}
    G_{N}(z) = \Omega_{N}\left(G^{(N)}_s(z)\right) \hspace{5pt}.
\end{equation}
In other words, we approximate $G(z)$ with another function $G_N(z)$, whose continued fraction coefficients agree with those of $G(z)$ up to and including $b^2_{N}$, but which are approximated with the sequence $b^2_{n,s}$ beyond that point:
\begin{equation}\label{eq:stitch}
    G_N(z) = \cfrac{1}{z-\cfrac{b_1^2}{{\cfrac{\ddots}{z-\cfrac{b^2_{N}}{z-\cfrac{\color{blue}b^2_{N+1,s}}{z- \cfrac{\color{blue}b^2_{N+2,s}}{\ddots}}}}}}} \hspace{5pt}.
\end{equation}
The aim of this discussion is to investigate the convergence of this approximation. Let  $\mathcal{L}_{N,s}$ denote the tridiagonal matrix with off-diagonal matrix elements $b_1,\ldots ,b_N,b_{N+1,s},b_{N+2,s}\ldots$. Expanding the resolvent Eq.~\eqref{G:resolvent} around this matrix gives an expression for the difference between the stitched and exact Green's function
\begin{align}
e_{N,s}(z) &= G_N(z) - G(z) \nonumber \\
           &= \langle O_0, (z - \mathcal{L})^{-1} (\mathcal{L}_{N,s} - \mathcal{L}) 
           (z - \mathcal{L}_{N,s})^{-1} O_0 \rangle.
\label{eqn: Error}
\end{align}
 An upper bound for the error when $\Im(z)\neq 0$ can be obtained by using the fact that $||(z-\mathcal{L})^{-1}||\leq 1/|\Im(z)|$ and $||\mathrm{J}||\leq 2\sup_{n}|b_{n}|$ holds for Jacobi operators $\mathrm{J}$ with vanishing diagonal elements \cite{koelink2001spectraltheoryspecialfunctions}. These bounds lead to 
\begin{equation}
|e_{N,s}(z)| \leq \frac{2}{|\Im(z)|^2} \sup_{n> N}|b_{n,s}-b_n|,
\label{Bound_finite_Imz}
\end{equation}
which implies that, in the worst case scenario, the rate of convergence is given by how quickly $|b_{N,s}-b_N|$ approaches zero. Hence, a sufficient condition for the stitching approximation to converge, $G_N(z) \to G(z)$, when $\Im(z) \neq 0$ is that the difference in the stitched and exact recurrence coefficients tends to zero: $|b_{N,s}-b_N| \to 0$.
\section{Zero-frequency approximation}
In the previous section, Eq.~\eqref{Bound_finite_Imz} shows that a sufficient condition for the convergence of the stitching approximation, when \( |\Im(z)| > 0 \), is that the stitched Lanczos coefficients become increasingly close to the exact ones, i.e., \( |b_{N,s} - b_N| \to 0 \) as \( N \to \infty \). However, it is unclear what happens in the low-frequency regime, as \( \Im(z) \to 0^- \), since Eq.~\eqref{Bound_finite_Imz} cannot be used to prove convergence in this limit. This regime is of particular interest in hydrodynamic studies because many key hydrodynamic quantities, such as diffusion constants, are extracted from the low-frequency behaviour of the system. Understanding this limit is therefore crucial for accurately determining diffusion constants, as we will discuss further in the next section.

The subtlety arises from an order of limits. If we keep $\Im(z)$ finite and assume that $|b_{n,s}-b_{n}| \to 0$ we can use Eq.~\eqref{Bound_finite_Imz} to conclude
\begin{equation}
   \lim_{z \to -\mathrm{i}0^{+}}\lim_{N \to \infty}G_N(z) = G(-\mathrm{i}0^{+}) = \mathrm{i}\pi \rho(0) \hspace{5pt},
   \label{Limit 1}
\end{equation}
where $\rho(0)$ is the spectral function at zero frequency. Ideally,  we would like to set $z \to -\mathrm{i}0^{+}$ first before taking $N \to \infty$ because this allows us to write an explicit expression for $G(-\mathrm{i}0^{+})$ in terms of the Lanczos coefficients as we will later discuss. Therefore, the crucial question is whether the following limits holds:
\begin{equation}
\lim_{N \to \infty} \lim_{z \to -\mathrm{i}0^{+}}G_N(z) \stackrel{?}{=} G(-\mathrm{i}0^{+})  \hspace{5pt}.
\label{Assumption}
\end{equation}
A priori, it is not guaranteed that we can swap the order of limits. 
To simplify the discussion, we specialize to chaotic Hamiltonians.
An important conjecture related to these systems is the Operator Growth Hypothesis (OGH) which states that the leading $n\to\infty$ asymptotic of the Lanczos coefficients generically saturates an upper bound determined by locality which depends on the dimension $d$ \cite{PhysRevX.9.041017}:
\begin{equation}
   b_n = 
   \begin{cases}
       \alpha \frac{n}{\log n} + \gamma + o(1), & d=1; \\ 
       \alpha n + \gamma + o(1), & d>1. 
    \end{cases}
    \label{eqn: OGH}
\end{equation}
To begin, we discuss the case $d > 1$ because we can perform the stitching approximation using the sequence $b_{n,s} = \alpha\sqrt{n(n-1+\eta)}$, which corresponds to the symmetric Meixner-Pollaczek polynomials \cite{PhysRevX.9.041017}. This sequence is well suited to deal with the case since, according to Eq.~\eqref{eqn: OGH}, we can tune the parameter $\alpha$ to ensure $b_n-b_{n,s} \to 0$; furthermore, we can also choose $\eta$ such that $\gamma$, in Eq.~\eqref{eqn: OGH}, is matched. On the other hand, in one dimension, to the best of our knowledge, no suitable solution is currently available exhibiting the required $n/\log{n}$ growth. Although the use of Meixner-Pollaczek stitching should give convergent results in $d>1$ for nonzero $\Im(z)$ (see bound Eq.~\eqref{Bound_finite_Imz}), we cannot prove the same statement for $d=1$. Nevertheless, following Ref.~\cite{PhysRevX.9.041017}, we proceed with this approach regardless in $d=1$, and starting from stronger assumptions we will analyse the rates of convergence in this case.

To make progress, we assume that the stitching approximation works in the zero-frequency limit, i.e. that Eq.~\eqref{Assumption} is valid, regardless of the dimension, which leads to the following assumption.
\begin{assumption}
The following limit holds:
\begin{equation}
    \lim_{N \to \infty} \lim_{z \to -\mathrm{i}0^{+}}G_N(z) = G(-\mathrm{i}0^{+}) \hspace{5pt},
\end{equation}
for the Meixner-Pollaczek stitching in $d\geq 1$, provided that $0 < |G(-\mathrm{i}0^{+})| < \infty$, and $b_N/b_{N,s} \to 1$.
\label{Convergence assumption}
\end{assumption}
The requirement $b_N/b_{N,s} \to 1$, will become clear in the next subsection, after we explicitly obtain a formula for $G(-\mathrm{i}0^{+})$.
The main implications of Assumption~\ref{Convergence assumption} are twofold. First, it gives us an explicit expression for $G(-\mathrm{i}0^{+})$ in terms of an infinite product of Lanczos coefficients. Second, it provides a new upper bound that, unlike that in Eq.~\eqref{Bound_finite_Imz}, is valid even in the zero-frequency limit.
With this assumption in hand, our task, in the next section, will be to understand how quickly the approximate Green's function $G_N$ converges to the true Green's function $G$ as $N$ increases.
\subsection{Infinite product}
An explicit formula for $G_N(-\mathrm{i}0^{+})$ can be found by iterating the M\"{o}bius map in Eq.~\eqref{eq:iterative} and setting $z \to -\mathrm{i}0^{+}$:
\begin{equation}
 G_N(-\mathrm{i}0^{+}) =
    \begin{cases}
    \frac{b_{N}}{b_{N,s}}\frac{G_s(-\mathrm{i}0^{+})}{\Pi_{N,s}}\Pi_{N}  & \text{$N$ even}  \\
    \frac{b_{N,s}}{b_{N}}\frac{G_s(-\mathrm{i}0^{+})}{\Pi_{N,s}}\Pi_{N}
    & \text{$N$ odd}
    \end{cases}
    \label{eqn: Zero-frequency stitching}
\end{equation}
where we have defined
\begin{equation}
    \Pi_{n} = \frac{1}{b_{n}}\prod_{k=1}^{\lfloor n/2\rfloor}\frac{b_{2k}^2}{b_{2k-1}^2} \hspace{5pt},
    \label{eqn: Product}
\end{equation}
and
\begin{equation}
    \Pi_{n,s} = \frac{1}{b_{n,s}}\prod_{k=1}^{\lfloor n/2\rfloor}\frac{b_{2k,s}^2}{b_{2k-1,s}^2} \hspace{5pt}.
\end{equation}
 We now make use of Assumption~\ref{Convergence assumption}, $G_N(-\mathrm{i}0^{+}) \to G(-\mathrm{i}0^{+})$, to derive a formula that links $G(-\mathrm{i}0^{+})$ to an infinite product involving the Lanczos coefficients. 
In order to do so, we first make a key observation regarding Eq.~\eqref{eqn: Zero-frequency stitching}: the ratio $G_s(-\mathrm{i}0^{+})/\Pi_{N,s}$ has the following limiting value
\begin{equation}
    \lim_{N \to \infty} \frac{G_s(-\mathrm{i}0^{+})}{\Pi_{N,s}} = \mathrm{i}
    \hspace{5pt}.
    \label{eqn: Known Limit}
\end{equation}  
This can be proven to be valid for the recurrence coefficients of the Meixner-Pollaczek polynomials \cite{sm}. Provided that Assumption~\ref{Convergence assumption} is valid, i.e. that $G_N(-\mathrm{i}0^{+}) \to G(-\mathrm{i}0^{+})$, we obtain the following limit
\begin{equation}
    \lim_{N \to \infty}\mathrm{i}\Pi_N = \lim_{N \to \infty}\frac{\mathrm{i}}{b_{N}}\prod_{k=1}^{\lfloor N/2\rfloor}\frac{b_{2k}^2}{b_{2k-1}^2}=G(-\mathrm{i}0^{+}) \hspace{5pt},
    \label{Conjecture}
\end{equation}
which links an infinite product of Lanczos coefficients to the Green's function at the origin. A similar formula appears in Ref.~\cite{PhysRevB.110.104413}. Motivated by the existence of an exact solution, which exhibits the correct asymptotic growth of the Lanczos coefficients—given by the Meixner–Pollaczek polynomials—we have focused our discussion on the OGH in $d > 1$. However, note that Assumption~\ref{Convergence assumption} also applies to $d=1$, so that Eq.~\eqref{Conjecture} also holds there. Moreover, from Eq.~\eqref{eqn: Zero-frequency stitching} we find that Meixner–Pollaczek stitching remains valid in $d=1$ provided the stitching is done in such as way as to guarantee  $\lim_{N\rightarrow\infty}b_{N,s}/b_N = 1$.
Strictly speaking, Equation~\eqref{Conjecture} was derived under the assumption that $G(-\mathrm{i}0^{+})$ is finite and nonzero; however, it may still hold even if this condition is not satisfied.

The infinite product in
Eq.~\eqref{Conjecture} will be used later, especially when we do not have a valid stitching sequence (e.g. in $d=1$). First, we introduce another upper-bound on the error. This bound serves two purposes, the first is to justify a formula which is used to find the rate of convergence of $G_N(-\mathrm{i}0^{+})$ presented in the next section, Eq.~\eqref{eqn: Asymptotics of stitch error}. The second is that it works even at zero frequency unlike the previous ones in Eq.~\eqref{Bound_finite_Imz}.
\subsection{Improved bound on the stitching error}
Before analysing the convergence rate of $G_N(-\mathrm{i}0^{+})$ in the next section, we first need to manipulate Eq.~\eqref{eqn: Error} into a more convenient form:
\begin{align}
e_{N,s}(-\mathrm{i}0^{+}) = \lim_{z \to -\mathrm{i}0^{+}}\sum_{n > N} &(b_n - b_{n,s}) \Bigl( C_n(z)C_{n-1}(z;N) \nonumber \\
& + C_{n-1}(z)C_n(z;N) \Bigr)
\label{eqn: Stitch_error} \hspace{5pt}.
\end{align}
Here, we have introduced the following matrix elements of the resolvents:
\begin{eqnarray}
C_{n}(z) &=& \langle O_0,(z-\mathcal{L})^{-1}O_n\rangle \nonumber \\
C_n(z;N)&=&  \langle O_0,(z-\mathcal{L}_{N,s})^{-1}O_n\rangle  \hspace{5pt}.
\end{eqnarray}
At this stage, we wish to interchange the summation and the limit in Eq.~\eqref{eqn: Stitch_error}, as this greatly simplifies the analysis of the rate of convergence (see Eq.~\eqref{eqn: Asymptotics of stitch error}). However, before proceeding, we must first establish a sufficient condition under which this interchange is valid. This analysis leads to a new upper bound, which, unlike that in Eq.~\eqref{Bound_finite_Imz}, holds for all frequencies along the negative imaginary axis:
\begin{equation}
|e_{N,s}(-\mathrm{i}y)|  \leq 2M\sum_{n > N}|b_{n,s}-b_n|b_{n}^{-1} \hspace{5pt}.
\label{eqn: zero-frequency bound}
\end{equation}

The discussion of this upper bound constitutes the content of the current subsection.
This upper bound is useful because, when finite, it provides sufficient conditions for the uniform convergence of the stitching approximation along the imaginary axis. Uniform convergence of the summation in  Eq.~\eqref{eqn: Stitch_error} is a strong condition because it justifies interchanging the summation and the limit. As a result, it simplifies the analysis in the next section, where we explicitly address the decay of the stitching error in the context of extracting the diffusion constant in chaotic systems. 
To infer uniform convergence we use the Weierstrass $M$-test, which implies that, if the summation on the right-hand side of Eq.~\eqref{eqn: zero-frequency bound} converges
\begin{equation}
\sum_{n >N}|b_{n,s}-b_n|b_{n}^{-1} < \infty \hspace{5pt},
\label{eqn: Uniform convergence}
\end{equation}
then the stitching error tends to zero uniformly along the imaginary axis:
\begin{equation}
    e_{N,s}(-\mathrm{i}y) \overset{\mathrm{uniform
 }}{\longrightarrow} 0 \hspace{5pt} \forall y \geq 0 \hspace{5pt}.
\end{equation}
Therefore, Eq.~\eqref{eqn: Uniform convergence} provides the sufficient condition we required.

Next, we briefly discuss the derivation of  Eq.~\eqref{eqn: zero-frequency bound}.
The task is to find an upper bound for Eq.~\eqref{eqn: Stitch_error} along the imaginary axis: $z=-\mathrm{i}y$ with $y\geq 0$. Specifically, we aim to find an upper bound for products of the form $C_n(z)C_{n-1}(z;N)$. To achieve this, we first derive a generic bound for expressions like $C_n(-\mathrm{i}y)$.
By the spectral theorem, these matrix elements are Cauchy-Stieltjes transforms \cite{simon2015operator}
\begin{equation}
C_n(z) = \langle O_n, (z-\mathcal{L})^{-1} O_0 \rangle = \int_{\mathbb{R}}\mathrm{d}\mu(x) \frac{p_n(x)}{z-x}   \hspace{5pt}, 
\end{equation}
where $d\mu(x)$ is the spectral measure of $\mathcal{L}$ and $p_n(x)$ the corresponding orthogonal polynomials. The spectral representation implies that the Cauchy transform satisfies the same three-term recurrence as the orthogonal polynomials, Eq.~\eqref{eqn: Three-term recurrence: polynomials}, but with different initial conditions: $C_0(z)=G(z)$ and $C_1(z) = (zG(z)-1)/b_1$.  
We restrict the analysis to the negative imaginary axis, $z=-\mathrm{i}y$ with $y\geq 0$. In this case, it can be shown that
\begin{equation}
    C_n(-\mathrm{i}y) = \mathrm{i}^{n+1}c_n(y) \hspace{5pt},
    \label{eqn: Phase}
\end{equation}
with $c_n(y)$ positive: $c_n(y) \geq 0$ \cite{sm}. The fact that $c_n(y)$ is positive allows us to find an upper bound directly from the three-term recurrence relation satisfied by $c_n(y)$:
\begin{equation}
    b_{n+1}c_{n+1}(y) = -y c_n(y) + b_n c_{n-1}(y) \hspace{5pt}.
\end{equation}
Neglecting the negative term $-yc_{n}(y)$ and iterating the resulting inequality, we obtain
\begin{eqnarray}
c_{2m+1}(y) &\leq&   c_1(y) \prod_{k=1}^{m
}\frac{b_{2k}}{b_{2k+1}}  \nonumber\\
c_{2m}(y) &\leq&   c_0(y) \prod_{k=1}^{m}\frac{b_{2k-1}}{b_{2k}} \hspace{5pt},
\label{eqn: Bound}
\end{eqnarray}
where the equality holds when $y=0$. Due to the generality of the derivation, a similar bound also applies to $C_n(z;N)$.
If Assumption~\ref{Convergence assumption} holds, we obtain the following bound on the product \cite{sm}:
\begin{equation}
|C_{n}(-\mathrm{i}y)C_{n+1}(-\mathrm{i}y;N)| \leq \frac{M}{b_{n+1}} \hspace{5pt}.
\label{eqn: Bound on product of Cauchy}
\end{equation}
Where $M$ is some positive constant, which does not depend on the stitching level $N$. In the special case $y=0$, the equality holds asymptotically for large $n$.
Using the inequality in Eq.~\eqref{eqn: Bound on product of Cauchy} and Eq.~\eqref{eqn: Stitch_error}, we obtain the improved upper-bound on the error presented in Eq.~\eqref{eqn: zero-frequency bound}.

Notice that, as anticipated, the upper-bound in Eq.~\eqref{eqn: zero-frequency bound} works even at zero frequency: $y \to 0^{+}$. Furthermore, it is useful because it allows us to formulate a sufficient condition for uniform convergence of the approximation. This allows us to interchange the limit $z \to -\mathrm{i}0^{+}$ and the summation in Eq.~\eqref{eqn: Stitch_error}, which significantly simplifies the analysis in the next section when we determine how fast the stitching error decays. Essentially, the reason for this simplification is that the three-term recurrence relation greatly simplifies when $z=0$. 

We conclude this discussion by deriving sufficient conditions for Eq.~\eqref{eqn: Uniform convergence} to hold provided that $b_n$ follows the OGH. We conclude that the rate at which the difference $b_n-b_{n,s}$ should decay in the marginal case is $1/\mathrm{poly}(\log n)$; results are summarised in Table~\ref{tab: Sufficient conditions}.

\begin{table}
    \centering
    \begin{tabular}{c@{\hskip 0.5in}c@{\hskip 0.5in}c}
        \toprule \\ & $ d=1 \hspace{5pt}$ & $d>1$ \\[0.1em]\\ \hline & \\ 
        $b_n$ \text{(OGH)} & $\alpha \frac{n}{\log n} +\mathcal{O}(1)$ & $\alpha n +\mathcal{O}(1) $  \\  [1em]
        $b_n-b_{n,s}$ &  $(\log n)^{-2-\epsilon}$ &  $(\log n)^{-1-\epsilon}$  \\ [0.01em] \\
        \botrule
    \end{tabular}
    \caption{
    The first row reminds the reader the asymptotic form of the Lanczos coefficient when the OGH holds, depending on the dimension $d$. 
    If the difference between the exact and the stitched coefficient, $b_n - b_{n,s}$, decays faster than what is listed in the second row ($\epsilon>0$), then it is sufficient to have  uniform convergence of the stitching approximation along the imaginary axis. 
    }
    \label{tab: Sufficient conditions}
\end{table}
\section{Diffusion constant}
In the previous section, we presented some assumptions that allow us to address the zero-frequency limit of the stitching approximation. These assumptions enable us to derive an infinite product linking the Lanczos coefficients to the spectral function at the origin and to establish a new upper bound for the error term that remains valid even at zero frequency.

In this section, we examine the rate of convergence of the stitching approximation. This is motivated by a practical problem in the study of chaotic Hamiltonians with diffusive transport: extracting the diffusion coefficient. In general, estimating the diffusion constant \( D \) in such systems is particularly challenging due to the rapid growth of entanglement, which complicates tensor network simulations \cite{PhysRevLett.111.127205}. However, the Lanczos algorithm offers a promising alternative for tackling this problem \cite{PhysRevB.110.104413}. In this section, we specialise to systems where the OGH is valid and we analyse how well appropriate stitching techniques perform. 
\subsection{Conductivity as a spectral function}
Before addressing the problem of determining the rate of convergence of the stitching approximation, we briefly establish the connection between the diffusion constant and the zero-frequency value of a specific spectral function.

The diffusion coefficient \( D \) is directly connected to the conductivity, \( \sigma(\omega) \), which is proportional to a spectral function. 
In general, the conductivity at inverse temperature $\beta$ is given by the Kubo formula
\begin{equation}
\sigma(\omega) = \beta \lim_{t \to \infty}\lim_{L \to \infty}\frac{1}{L}\int_{0}^{t}\mathrm{d}t' e^{-\mathrm{i} \omega t'} K_{JJ}(t')
\end{equation}
where $L$ is the system's size, $J$ is the current, 
\begin{equation}
K_{AB} = \frac{1}{\beta}\int_{0}^{\beta}\mathrm{d}\lambda \langle AB(t+\mathrm{i}\lambda)\rangle_{\beta} 
\end{equation}
is the Kubo correlator, and $\langle \,\cdot  \,\rangle_{\beta} = \mathrm{tr}[e^{-\beta H}\,\cdot \,]/\mathrm{tr}[e^{-\beta H}]$ is the thermal expectation value \cite{RevModPhys.93.025003}.
In the absence of a Drude weight, the conductivity at zero frequency is proportional to the diffusion constant 
\begin{equation}
    D = \frac{\sigma(0)}{\chi}
\end{equation}
where $\chi = \lim_{L \to \infty}\beta(\langle Q^2\rangle_{\beta} -\langle Q\rangle_{\beta}^2) / L$ is the static susceptibility and $Q$ the conserved charge. We restrict our analysis to the infinite temperature case for which the diffusion constant is given by
\begin{equation}
D =  \frac{1}{\langle q_0,q_0 \rangle}\lim_{\omega \to 0}\lim_{t \to \infty}\lim_{L \to \infty}\frac{1}{L}\int_{0}^{t}\mathrm{d}t' e^{-\mathrm{i} \omega t'} \langle J,J(t')\rangle
\end{equation}
where $q$ is the charge density and $\langle \, \cdot \,, \, \cdot \, \rangle$ is the Hilbert–Schmidt inner product. Therefore, we conclude that $D$ is proportional to the spectral function at zero frequency associated with the current operator $J$. 

This connection suggests initialising the Lanczos algorithm with $O = J/||J||$, which can be achieved using translational invariance \cite{PhysRevX.9.041017}. Subsequently, a suitable ``stitching" procedure can be used to approximate the diffusion constant using the first $N$ Lanczos coefficients. 
To emphasise this connection, we re-write Eq.~\eqref{Conjecture} to better suit the current context as
\begin{equation}
D  = \frac{\langle j_{0}, j_{0}\rangle}{\langle q_{0}, q_{0}\rangle}\lim_{N \to \infty}\frac{1}{b_{N}}\prod_{k=1}^{\lfloor \frac{N}{2} \rfloor}\frac{b_{2k}^2}{b_{2k-1}^2} \equiv \lim_{N \to \infty}D_N   \hspace{5pt} \label{eqn: Diffusion constant};
\end{equation}
this formula also appears in Ref.~\cite{PhysRevB.110.104413}.

Before analyzing the rate of convergence of the stitching approximation, it is natural to ask: what condition on the asymptotic form of \( b_n \) ensures that the diffusion constant \( D \) remains finite?  

This is an important preliminary step because, without ensuring that \( D \) remains finite, analyzing the rate of convergence would be an ill-posed question. 
In other words, we seek a criterion for the associated product to converge. To address this, we assume that the Lanczos coefficients \( b_n \) exhibit subleading oscillations, which are expected to influence the convergence behaviour owing to the ratio in Eq.~\eqref{eqn: Diffusion constant}.  
This assumption is further supported by numerical findings and more rigorous results obtained using the steepest descent method to analyse the Riemann–Hilbert problem associated with orthogonal polynomials. In particular, Ref.~\cite{10.1155/S1073792899000161} provides an example where the Lanczos coefficients grow linearly while exhibiting a slowly decaying even/odd staggered correction. Moreover, this example features a spectral function that is non-analytic at the origin, suggesting a connection between spectral singularities and staggered corrections, which we will discuss in the next section.  
Additionally, Ref.~\cite{10.1007/978-3-642-82444-9_3, VANLESSEN2003198}, despite focusing on bounded Lanczos coefficients that do not follow the OGH, demonstrates that algebraic singularities in the spectral functions can induce oscillatory corrections in the Lanczos coefficients.  
With these considerations in mind, we focus on the simplest form of oscillatory behaviour, namely even/odd staggering.

Hence, let $b_n = f_n + (-1)^n s_n$, where we have separated the staggered, $(-1)^ns_n$, and unstaggered part, $f_n$, in the asymptotic expansion of $b_n$. In order to ensure that $b_n$ is positive we require $s_n/f_n \to 0$. Since we are interested in chaotic Hamiltonians, where $f_n \propto n$ or $f_n \propto n/\log n$, we restrict $f_n$ such that $f_n \to \infty$. Analysing the asymptotic expansion of the product in Eq.~\eqref{eqn: Diffusion constant}, we obtain the following criterion for convergence \cite{sm}
\begin{equation}
  \left|\lim_{n \to \infty} \int \mathrm{d}n \frac{s_{n}}{f_{n}}\right| <\infty \implies 0<\rho(0)<\infty  \label{eqn: Criterion} \hspace{5pt},
\end{equation}
which implies that slowly decaying even/odd staggered terms are sufficient for this product to diverge/vanish. For simplicity, we express this criterion using an integral, assuming that \(s_n\) and \(f_n\) are continuous in \(n\); however, an equivalent formulation using summation is also possible. As evidenced by Eq.~\eqref{eqn: Criterion}, a sufficient condition to ensure a finite diffusion constant, for systems obeying the OGH, is that the staggering decays at least as fast as a power law: $s_n <n^{-\epsilon}$ for some positive $\epsilon$. 
\subsection{Convergence rate}
Having established a criterion for the finiteness of the diffusion constant, we now turn our attention to how different stitching procedures, as influenced by the dimension $d$, affect the rate of convergence. As we have discussed, the OGH predicts two distinct asymptotic behaviours based on the dimension, which influences the type of stitching procedure required. In dimensions greater than one ($d > 1$), there exists an exact solution that exhibits the correct asymptotic behaviour of the Lanczos coefficients. However, in one dimension, $d = 1$, no known solution  exhibits the required $n / \log n$ growth.
However, the content of Assumption~\ref{Convergence assumption} is that the MP stitching method converges to the correct solution in this case as well, although as we will see, the predicted convergence is usually slower than in $d>1$.

We first focus on the case $d > 1$. We perform the stitching approximation with the sequence $b_{n,s} = \alpha\sqrt{n(n-1+\eta)}$, which corresponds to the symmetric Meixner-Pollaczek polynomials \cite{PhysRevX.9.041017}. The goal is to determine the decay of $e_{N,s}(-\mathrm{i}0^{+})$, which reveals how quickly the approximation to the diffusion constant converges. Recall from Eq.~\eqref{eqn: Stitch_error}, that there is an exact expression for the stitching error:
\begin{align}
e_{N,s}(z) = \sum_{n > N} (b_n - b_{n,s}) \Bigl( & C_n(z)C_{n-1}(z;N) \nonumber \\
& + C_{n-1}(z)C_n(z;N) \Bigr) \hspace{5pt}.
\end{align}
For the sake of clarity in the discussion, we specialize to the case $|b_{n,s}-b_n| = \mathcal{O}\left(n^{-a}\right)$. We can then safely interchange the limit \( z \to -\mathrm{i}0^{+} \) and the summation, as uniform convergence is ensured by the results of the previous section. Using  Eq.~\eqref{eqn: Phase} and Eq.~\eqref{eqn: Bound on product of Cauchy} in the special case $y=0$, we obtain the following asymptotic expression valid at large $N$
\begin{equation}
e_{N,s}(-\mathrm{i}0^{+}) =  c\sum_{n>N} (b_n - b_{n,s})\frac{(-1)^n}{b_n} \hspace{5pt},
\label{eqn: Asymptotics of stitch error}
\end{equation}
where $c$ is a constant.
Therefore, we conclude that the decay of the stitching error at level \( N \), \( e_{N,s}(-\mathrm{i}0^{+}) \), depends on the presence of staggered terms in the difference $b_n-b_{n,s}$. We identify two cases:
\begin{enumerate}
    \item  $b_n-b_{n,s} = (-1)^n\mathcal{O}\left( n^{-a}\right)$ 
\begin{equation}
    e_{N,s}(-\mathrm{i}0^{+}) = \mathcal{O}(N^{-a}) \hspace{10pt}  \hspace{5pt},
    \label{eqn: Relevant staggered case}
\end{equation}
\item $b_n-b_{n,s} =  \mathcal{O}(n^{-a})$, and the next staggered term decays faster than $(-1)^{n}n^{-a-1}$
\begin{equation}
    e_{N,s}(-\mathrm{i}0^{+}) = \mathcal{O}(N^{-a-1}) \hspace{10pt}  \hspace{5pt}.
     \label{eqn: Irrelevant staggered case}
\end{equation}
\end{enumerate}
The conclusion is that, in general,  staggered terms lead to a slower decay rate compared to unstaggered terms. 

We now turn to the one-dimensional case, $d = 1$. In this setting, an exact solution with the correct leading-order asymptotic behaviour of $n / \log n$ is not available.
A proposed strategy is to approximate the diffusion constant by using a partial product $D_N$, as defined in Eq.~\eqref{eqn: Diffusion constant}, which includes only the first $N$ recurrence coefficients. This approximation converges to the right limit by Assumption~\ref{Convergence assumption}. 
This approach is equivalent to stitching a ``constant" sequence, $b_{n,s} = b_N$ for all $n > N$. For this reason, we refer to this method as ``constant stitching." By analysing the asymptotics of $D_N$, it can be shown that the rate of convergence is given by~\cite{sm}:
\begin{equation}
    D - D_N \propto
    \begin{cases}
        \max \left(\int \mathrm{d}N \frac{s_N}{N/\log N}, N^{-1}\right) & d = 1, \\
        \max \left(\int \mathrm{d}N \frac{s_N}{N}, N^{-1}\right) & d > 1.
    \end{cases}
    \label{eqn: Convergence of constant stitching}
\end{equation}
We conclude that, in the optimal case, constant stitching converges as $N^{-1}$. However, in the presence of slowly decaying staggered terms, the rate of convergence can become arbitrarily slow. For example, consider the extreme case where $s_N \propto (\log N (\log \log N)^2)^{-1}$ for $d > 1$. In this scenario, the convergence rate becomes exceptionally slow, with $D - D_N \propto 1/(\log \log N)$.

Having discussed the rate of convergence for both constant stitching and Meixner-Pollaczeck stitching, a natural question arises: when is the latter preferable over the former? To answer this, we now consider a concrete example that highlights the situations in which the Meixner-Pollaczeck stitching approximation is more effective than constant stitching. 

Consider a generic Green's function in $d>1$ obtained from $b_n = \alpha n +\gamma + (-1)^ns_n + d_{n}$ where $d_n = o(1)$ does not contain any staggered terms in its asymptotic expansion. We stitch the sequence associated to Meixner-Pollaczeck given by $b_{n,s} = \alpha \sqrt{n(n-1+\eta)} = \alpha n+ \gamma + \delta/n + o(n^{-1})$, with $\gamma = \alpha(\eta-1)/2$ and $\delta = \alpha(\eta-1)^2/8$. Assume that the two coefficients $b_n$ and $b_{n,s}$ match up to and including $\mathcal{O}(1)$. This can be usually done since $b_{n,s}$ contains a free parameter $\eta$ that controls the $\mathcal{O}(1)$ term. Furthermore, consider $s_n$ such that $D$ is finite; i.e. Eq.~\eqref{eqn: Criterion} holds. 

We identify two cases for the rate of convergence, based on the relevance of the staggered correction, which are summarised in Table~\ref{tab: Convergence}.
\begin{table}
    \centering
    \footnotesize
    \scalebox{0.97}{
    \begin{tabular}{ccc}
        \toprule \\ [-0.7em]
         & \textit{Relevant} &\textit{Irrelevant} \\
         & \text{\( n^{-1} \ll s_{n} \ll n^{-\epsilon} \)} & $s_{n} = o(n^{-1})$ \\[0.2em] \hline & \\[0.1em] 
        \text{Error of} & $s_N$ & $N^{-1}$ \\
        \text{constant stitching}& \\
        [1em]
        \text{Error of MP} & $s_N$ & $\max(N^{-2}, s_N, N^{-1}d_N)$\\ \text{stitching}& \\ [0.01em] \\
        \botrule
    \end{tabular}}
    
    \caption{Convergence rates at level $N$ for constant stitching and for the stitching approximation using the recurrence coefficients of the Meixner–Pollaczek (MP) polynomials, given by \( b_{n,s} = \alpha \sqrt{n(n-1+\eta)}. \) The aim is to approximate a Green's function at zero frequency, obtained with \( b_n = \alpha n + \gamma + (-1)^n s_n + d_n, \) where \( s_n \) is the staggering part and \( d_n = o(1) \). When staggering is relevant, constant and MP stitching have the same rate of convergence; whereas, when staggering is irrelevant, MP stitching converges faster. As a side note, we use the symbol \( \ll \) in an asymptotic sense.}
    \label{tab: Convergence}
\end{table}
When the staggered term decays sufficiently slowly (relevant staggering), there is no advantage to using the Meixner–Pollaczek stitching, as a partial product (constant stitching) achieves the same rate of convergence. However, when the decay is fast enough (irrelevant staggering), the Meixner–Pollaczek stitching improves the rate of convergence. 

We conclude this section by outlining the key results. In $d > 1$, the stitching approximation is useful only when staggering is irrelevant, as defined in Table~\ref{tab: Convergence}; otherwise, it offers no advantage over constant stitching. In contrast, in $d = 1$, the simplest method for extracting the diffusion constant is constant stitching, due to the lack of an exact solution exhibiting the correct asymptotic growth, $n/\log n$, of the Lanczos coefficients. The convergence rate of this method, in the best-case scenario, is $N^{-1}$, as presented in Eq.~\eqref{eqn: Convergence of constant stitching}. The key conclusion of this analysis is that, in both cases, slowly decaying staggered terms significantly hinder convergence.
\section{Smoothness of the spectral function at the origin, and the resources required for estimating \texorpdfstring{$D$}{D}}
\label{Subsection: Smoothness of the spectral function at the origin, and the resources required for estimating $D$}
So far, we have discussed the importance of subleading staggered terms in determining the rate of convergence of the stitching approximation. However, the previous section does not address the fundamental question of when such subleading terms arise. Here we connect these staggered terms to non-analyticities in the spectral function at the origin. This insight is based on the following theorem by Magnus \cite{10.1007/978-3-642-82444-9_3}. Consider a symmetric spectral function of a self-adjoint operator that has compact support such that at the origin it has a power-law of the form $\rho(x) \propto |x|^{\alpha}$, then the asymptotic of the Lanczos coefficients is
\begin{equation}
    b_n = b_{\infty}\Bigl(1-(-1)^{n}\frac{\alpha}{2n} + o(n^{-1})\Bigl) \hspace{5pt}.
\end{equation}
Even though this theorem does not directly apply to the spectral functions of chaotic many-body systems, which are supported on the entire real line, it establishes a connection between staggered subleading terms and smoothness at the origin.

To further support this connection, consider that an $\mathcal{O}(1)$ staggering is sufficient to guarantee the presence of a Dirac delta function at the origin in the spectral measure of models with Lanczos coefficients obeying the OGH. This is a characteristic feature of many-body localisation (MBL) \cite{PhysRevB.93.134206,10.21468/SciPostPhys.13.2.037}, as it implies that auto-correlation functions do not decay in the long-time limit. This $\mathcal{O}(1)$ staggered term was detected numerically in the quantum Ising model with a longitudinal field and random transverse fields when the algorithm is initialised with a Pauli $Z$ operator \cite{10.21468/SciPostPhys.13.2.037}. We provide a short analytical argument explaining why this staggering leads to such behaviour.
A necessary and sufficient condition for the existence of a Dirac delta function at the origin is the convergence of the following sum of orthonormal polynomials \cite{Ismail_2005}:
\begin{equation}
    \sum_{n \geq 0} p_n^2(0) = S < \infty \hspace{5pt}.
    \label{eqn: Dirac}
\end{equation}
The weight associated to such a Dirac delta function is then $1/S$.
In this brief discussion, we specialise to $d=1$, since MBL is expected to be unstable in $d>1$ \cite{PhysRevB.95.155129}.
In view of this, consider the following asymptotic expansion of the Lanczos coefficients $b_n = n/\log n + s(-1)^n + o(1)$, with $s >0$. Using the asymptotic expansion of the infinite product $\Pi_n$ \cite{sm}, and its relation to the orthogonal polynomials given by
\begin{equation}
    \Pi_{2n} = \frac{1}{b_{2n}p^2_{2n}(0)} \hspace{5pt},
    \label{eqn: Product and even polynomials}
\end{equation}
it can be verified that $ p_n^2(0) = \mathcal{O}(n^{-1-s(\log n)} \log n) $. Therefore, the summation in Eq.~\eqref{eqn: Dirac} converges if the OGH holds and there is $\mathcal{O}(1)$ staggering: a signal of many-body localisation. Notice how it is important to enforce $s>0$ which ensures that $p^2_n(0)$ decays fast enough for the sum to converge. On the other hand, if $s<0$ the product $\Pi_n$ tends to zero and we expect, by Eq.~\eqref{Conjecture}, that the spectral function is zero at the origin: $\rho(0) = 0$.

\subsection{Criterion on the decay of the staggered terms}
This analysis suggests that slowly decaying staggered terms are associated either with non-analyticity of the spectral function or with its vanishing at the origin. We further support this claim with a non-rigorous calculation~\cite{sm}.  
In particular, we investigate how the differentiability of $G(z)$ at the origin affects the decay of the staggered terms in the Lanczos coefficients.  
This analysis is motivated by the fact that the conductivity is typically not infinitely differentiable at the origin, which, in turn, leads to a power-law decay of the current-current correlator in time~\cite{PhysRevB.73.035113}.

This calculation is performed using the following approximation of the iterated M\"{o}bius map:
\begin{equation}
    G(z) = \Omega_n\left(G^{(n)}(z)\right) \approx \Omega_n\left(\mathrm{i}/b_n\right) \hspace{5pt},
\end{equation}
which is expressed in terms of the \(n\)-th level Green's function, as defined in Eq.~\eqref{eqn: nth level Green's function}. While we cannot rigorously prove the validity of this approximation, we can justify why it is reasonable to replace \( G^{(n)}(z) \) with \( \mathrm{i}/b_n \). Specifically, it can be shown that if \( \lim_{n \to \infty} b_n G^{(n)}(z) \) exists, the only consistent solution is \cite{sm}: 
\begin{equation}
 \lim_{n \to \infty} b_n G^{(n)}(z) = \mathrm{i} \hspace{20pt} \Im(z) < 0 \hspace{5pt}.
\end{equation}
Furthermore, this limit can be extended to the case \( z \to -\mathrm{i}0^{+} \), provided that Assumption~\ref{Convergence assumption} holds. In fact, it can be shown that the limit of the infinite product in Eq.~\eqref{Conjecture} is equivalent to \cite{sm}: 
\begin{equation} 
\lim_{n \to \infty} b_n G^{(n)}(-\mathrm{i}0^{+}) = \mathrm{i} \hspace{5pt}.
\end{equation}

Therefore, we assume that $b_nG^{(n)}(z) \to \mathrm{i}$ holds when $\Im(z)\leq 0$ and then demand $G(z)$ to be $(k-1)$-differentiable but not $k$-differentiable at the origin.
In other words, we impose the condition that the $k$-th derivative diverges at the origin, while all the previous derivatives remain finite. Note that, in this analysis, the origin is approached from the lower half of the complex plane.
This analysis leads to the following criterion for the staggered part $s_n$ of the Lanczos coefficients obeying the OGH in $d>1$ (i.e. $b_n \propto n$). If the $k$-th derivative diverges but the $(k-1)$-th does not, then asymptotically these two conditions must hold simultaneously \cite{sm}:
\begin{eqnarray}   
   &\phantom{=}&  \lim_{n \to \infty} (\log n)^{k-1} \int_{\mathcal{O}(1)}^{n}\frac{\mathrm{d}n_1}{n_1}\int_{\mathcal{O}(1)}^{n_1}\frac{\mathrm{d}n_2}{n_2} s_{n_2} = \infty   \nonumber\\  
      \nonumber\\  
    &\phantom{=}& \lim_{n \to \infty}(\log n)^{k-2} \int_{\mathcal{O}(1)}^{n}\frac{\mathrm{d}n_1}{n_1}\int_{\mathcal{O}(1)}^{n_1}\frac{\mathrm{d}n_2}{n_2}s_{n_2}  < \infty \hspace{5pt}.
    \label{eqn: k-1-differentiability}
\end{eqnarray}
The integral should be computed ignoring any additive constant that arises from the lower limits of integrations.
These conditions are valid when $k\geq 2$. On the other hand, when $k=1$, we only require the first equation in Eq.~\eqref{eqn: k-1-differentiability} to hold.
The condition in Eq.~\eqref{eqn: k-1-differentiability} is met if the staggered term $s_n$ decays as $1/\mathrm{poly}(\log{n})$. A similar condition can then be found for Lanczos coefficients obeying the OGH in $d=1$ (see Eq.~\eqref{eqn: d=1, k-1-differentiability}). 

We summarise the criterion in Table~\ref{tab: Analyticity}.
\begin{table}
   \centering
   \scalebox{0.87}{\begin{tabular}{c@{\hskip 0.8cm}c@{\hskip 0.8cm}c}
       \toprule \\ [-0.7em]
        \text{Order of first }& $d=1$ & $d>1$\\ \text{divergent derivative}&  & \\\hline & \\[0.1em] 
        $k=1$ & $2<a\leq4$ & $1<a\leq2$\\
       $k=2$ & $4<a<6$ &  $2<a<3
       $ \\
       $k>2$  & $2k \leq a<2(k+1)$ &  $k\leq a<k+1$
       \\ [0.01em] \\
       \botrule
   \end{tabular}}
   \caption{Range of the decay parameter $a$ of the staggered term $s_n = (\log n)^{-a}$, depending on the dimension $d$, such that the $k$-th derivative is the first divergent derivative of the spectral function at the origin.}
   
   \label{tab: Analyticity}
\end{table}
Notice that this criterion has the desirable property that requiring a smoother spectral function (larger $k$) puts a more stringent condition on the rate of decay of the staggering. Furthermore, this analysis aligns with the well-known case of the Freud weight, where the spectral function is given by $\rho(x) = \frac{\pi}{2} e^{-\pi |x|}$. In this case, the Green's function has the following small-$z$ expansion:
\[
G(z) = \frac{\mathrm{i} \pi^2}{2} - \pi^2 z \log z + \mathcal{O}(z),
\]
which exhibits a divergent derivative ($k = 1$) at the origin. 
It has been proved that the Lanczos coefficients have the asymptotic form:
\[
b_n = \frac{n}{2} + \frac{(-1)^n}{(2\log n)^2} + o((\log n)^{-2}),
\]
as shown in Ref.~\cite{10.1155/S1073792899000161}.
Using the first line in Eq.~\eqref{eqn: k-1-differentiability}, we can see that if we substitute $s_n \propto (\log n)^{-2}$ and set $k=1$ we get that the expression diverges, which signals that $G(z)$ is not differentiable at the origin.

\subsection{Complexity of estimating the diffusion constant}
\label{Subsection: Complexity of estimating the diffusion constant}
We can now estimate the algorithm's complexity, assuming that we want to calculate the value of the spectral function at the origin with precision \( \epsilon \). If the error scales polynomially as \( N^{-\xi} \), then approximately \( N = \mathcal{O}(\epsilon^{-1/\xi}) \) iterations are needed to achieve this precision. On the other hand, if the error scales as \( (\log N)^{-\xi} \), then to achieve \( \epsilon \) precision, roughly \( N = \mathcal{O}(\exp(\epsilon^{-1/\xi})) \) iterations are required.  

In the context of estimating the diffusion constant, we need to consider the conductivity, which is generally non-analytic at zero frequency due to long-time hydrodynamic tails.
For instance, when there are two conserved quantities in $d=1$ and transport is diffusive it is expected that $\langle J,J(t) \rangle \propto t^{-3/2}$, which implies that the conductivity is finite but not differentiable at the origin \cite{PhysRevB.73.035113}. This means that $G(-\mathrm{i}0^{+})$ is finite whereas $G^{(1)}(-\mathrm{i}0^{+})$ diverges ($k=1$).
In this case, we expect the staggered term to decay as $s_{n} \propto (\log n)^{-a}$ with $2<a\leq 4$, according to  Table \ref{tab: Analyticity}. Therefore, using Eq.~\eqref{eqn: Convergence of constant stitching}, we expect the convergence rate to be $1/\mathcal{O}((\log{N})^{a-2})$. On the other hand, if we have only one conserved quantity (usually energy density) and transport is diffusive and the Hamiltonian is translation invariant and parity symmetric we expect the current-current correlator to decay as $\langle J,J(t) \rangle \propto t^{-4}$ \cite{PhysRevB.110.134308}. This suggests that  the conductivity is twice but not three times differentiable at the origin ($k=3$). Consequently, in this case, we predict the decay of the staggered part to be $s_n \propto (\log n)^{-a}$ with $6\leq a<8$, as shown in Table~\ref{tab: Analyticity}. This implies, from Eq.~\eqref{eqn: Convergence of constant stitching}, that in this case, the convergence rate of constant stitching lies between $1/\mathcal{O}((\log{N})^{4})$ and $1/\mathcal{O}((\log{N})^{8})$.

Thus, for the task of estimating diffusion constants, an accuracy $\epsilon$ approximation appears to require $N=\exp(\mathrm{poly}(1/\epsilon))$ where the form of the polynomial depends on the exponent in the decay of $\langle J,J(t) \rangle$. How does the computational cost (CPU/memory) of an $\epsilon$-approximation scale with $\epsilon$? That will depend on how difficult it is to compute the first $N$ Lanczos coefficients; naively it requires $\exp(\mathcal{O}(N))$ CPU-time, but recent work using matrix product operators suggests $\mathrm{poly}(\mathcal{O}(N))$ as a possibility \cite{PhysRevB.102.035147}, which would mean a net CPU time scaling as $\exp(\mathrm{poly}(1/\epsilon))$.
\section{Numerical results}
\begin{figure}
    \centering
    \includegraphics[scale=0.26]{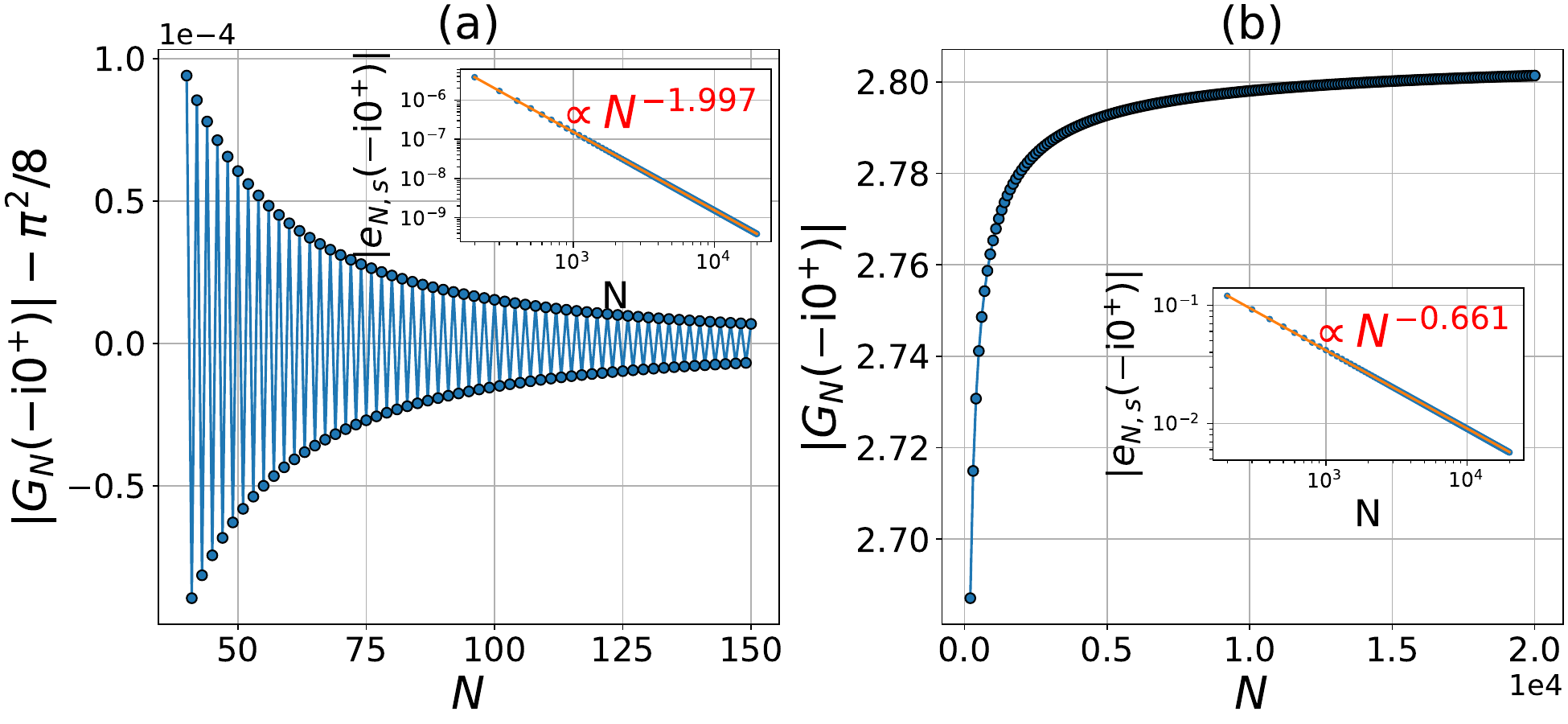}
    \caption{Stitching approximation for irrelevant  (a) and relevant (b) cases. For the irrelevant case there is an exact expression $|G(-\mathrm{i}0^{+})| = \pi^2/8$. However, the relevant case does not have a closed expression; therefore, we approximate $G(-\mathrm{i}0^{+})$ using a partial product (constant stitching) in Eq.~\eqref{eqn: Product} which yields $|G(-\mathrm{i}0^{+})| \approx 2.8071$. The error is predicted to scale as $N^{-2}$ for the irrelevant case and as $N^{-2/3}$ for the relevant case; these predictions agree with the fitting curves plotted in the insets.}
    \label{fig: Convergence}
\end{figure}
We now support the previous analytic predictions with numerical calculations. We consider two cases motivated by the two possible asymptotics of the Lanczos coefficients according to the OGH.
For linear Lanczos growth (believed to apply in $d>1$ chaotic systems), we perform the stitching approximation using the exact solution from the Meixner–Pollaczek polynomials and we support the claim that this approximation is beneficial only if staggering is irrelevant (in the sense of Table~\ref{tab: Convergence}).
Lastly, we focus on an explicit $d=1$ spin chain, computing the energy diffusion constant in the chaotic Ising model using the infinite product formula Eq.~\eqref{eqn: Diffusion constant}. Our result agrees well with a previous prediction from the original operator growth hypothesis work \cite{PhysRevX.9.041017}, which uses a more elaborate extrapolation procedure. 
\begin{figure}
        \centering
        \includegraphics[scale=0.262]{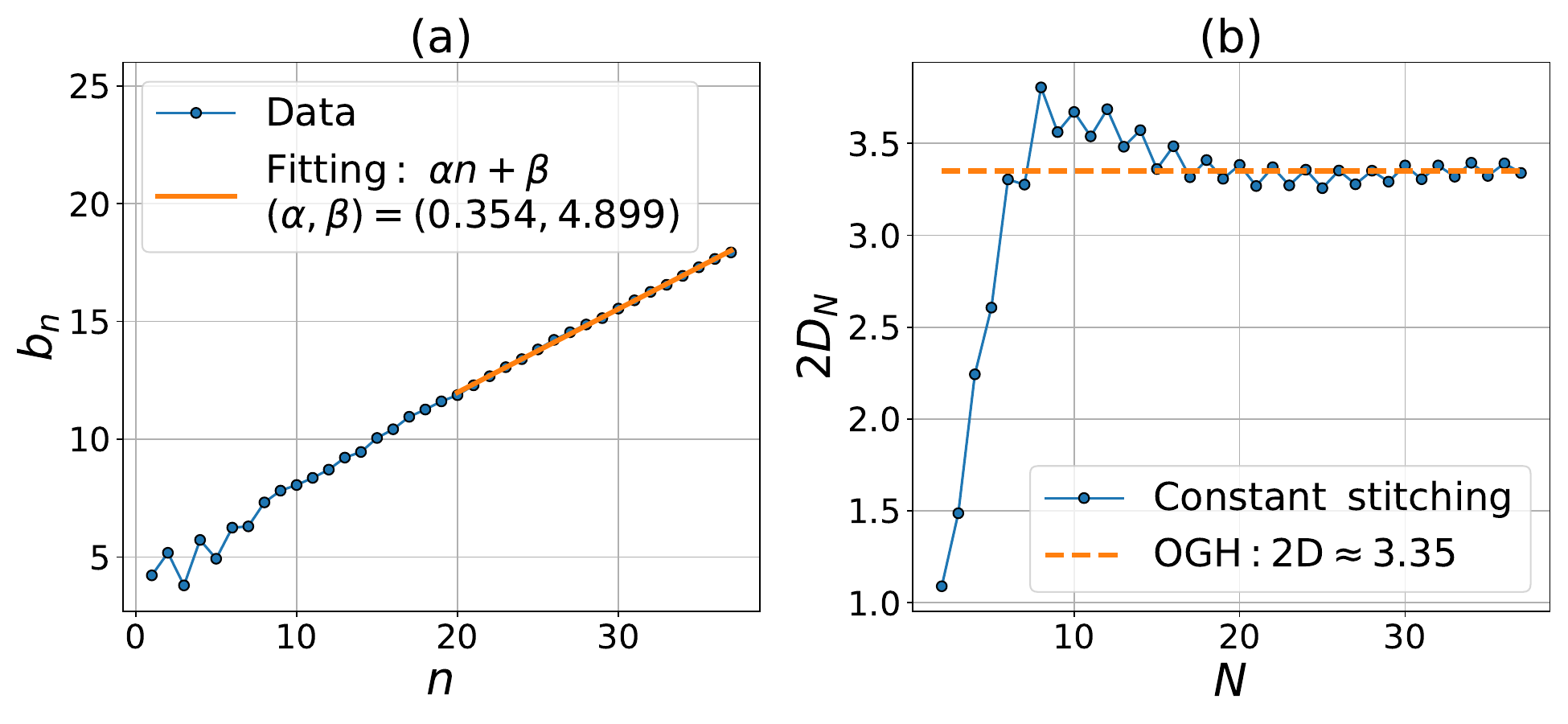}
    \caption{
    (a) Lanczos coefficients obtained by initialising the algorithm with a normalised current operator. The coefficients are expected to grow as $n/\log(n)$. However, resolving the logarithm and the sublinear corrections is challenging. (b) Diffusion constant as a function of the number of Lanczos coefficients $n$. We have rescaled it by a factor of two to compare it with the result quoted in Ref.~\cite{PhysRevX.9.041017} which is the dashed line $2D \approx 3.35$.}
    \label{fig: Diffusion constant}
\end{figure}
Let us begin by considering the following toy models with linear Lanczos growth and subleading terms which are irrelevant/relevant respectively (Table~\ref{tab: Convergence}):
\begin{enumerate}[label=\alph*)]
    \item Irrelevant staggering: $b_n = \frac{n^2}{\sqrt{n^2-1/4}} = n + \frac{1}{8}n^{-1} + o(n^{-1})$
    \item Relevant staggering: $b_n = n+ 1+ \frac{(-1)^n}{2}n^{-2/3}$.
\end{enumerate}
The first example has an exact solution for the correlation function $C(t) = 2t/\sinh(2t)$, whose Green's function has a known continued fraction expansion \footnote{See Eq. (94.13) p.373 in Ref.~\cite{wall2018analytic}}. On the other hand, the second case does not have a closed form.
The stitching is performed using the Meixner-Pollaczeck recurrence coefficients given by $b_{n,s} = \alpha \sqrt{n(n-1+\eta)} = \alpha n+ \gamma + \delta/n + o(n^{-1})$, with $\gamma = \alpha(\eta-1)/2$ and $\delta = \alpha(\eta-1)^2/8$. For the first case (irrelevant staggering), we use $(\alpha,\eta)= (1,1)$ so that the two sequences are matched up to and including $\mathcal{O}(1)$. Similarly, for the second case (relevant staggering) we use $(\alpha,\eta)=(1,3)$.
In the first case, our theory predicts error at the origin to scale as $ e_{N,s}(-\mathrm{i}0^{+})  \propto N^{-2}$ and $D-D_N \propto N^{-1}$ (irrelevant staggering) based on Eq.~\eqref{eqn: Irrelevant staggered case} and Eq.~\eqref{eqn: Convergence of constant stitching}. For the second case, we expect $ e_{N,s}(-i0^{+}) \propto D-D_N \propto N^{-2/3}$ (relevant staggering) according to Eq.~\eqref{eqn: Relevant staggered case} and Eq.~\eqref{eqn: Convergence of constant stitching}. These expectations are confirmed by numerically fitting the errors as show in Fig.~\ref{fig: Convergence}.

We next examine the case $d=1$. The aim is to approximate the diffusion constant for the mixed field Ising model $H = \sum_{i}\sigma^{x}_{i}\sigma^{x}_{i+1} + g_z\sigma^{z}_{i}+g_z\sigma^{x}_{i}$ with $g_z = -1.05$ and $g_x = 0.5$. In this regime, the model is expected to be chaotic and therefore energy density is diffusive \cite{PhysRevX.9.041017}. Unlike Ref.~\cite{PhysRevX.9.041017}, we use an infinite product formula  Eq.~\eqref{eqn: Diffusion constant} to approximate the diffusion constant; the validity of this formula follows from Assumption~\ref{Convergence assumption}. The results are presented in Fig.~\ref{fig: Diffusion constant}, and agree well with Ref. \cite{PhysRevX.9.041017}, despite the present method being significantly simpler to implement. Determining the exact rate of convergence is complicated because the subleading behaviour of the Lanczos coefficients is hard to extract, but we conjecture that it goes as $1/\mathrm{poly}\log(N)$ (see Subsection~\ref{Subsection: Smoothness of the spectral function at the origin, and the resources required for estimating $D$}).
The above numerical results corroborate the theory of errors presented in the previous section. In particular, they substantiate the claimed relationship between convergence rate and the subleading corrections to the operator growth hypothesis. 
\section{Discussion}
We have presented a procedure for computing the rate of convergence of the stitching approximation, which is used to estimate Green's functions. This approximation is based on completing a sequence of Lanczos coefficients with a sequence for which an exact solution to the Green's function is known $\{b_n\}_{n\leq N} \cup \{b_{n,s}\}_{n > N}$. There are different methods for completing such a sequence; ``MP" involves stitching using the Meixner-Pollaczek sequence of Lanczos coefficients for $b_{n,s}$.  ``Constant stitching" involves setting all Lanczos coefficients for $n>N$ to be a constant $b_{n>N,s}=b_N$ (the last calculated Lanczos coefficient). 

First, we derived an upper bound on the error for \(\Im(z) \neq 0\), as given in Eq.~\eqref{Bound_finite_Imz}. This analysis shows that the rate of convergence is governed by the decay of the difference between the exact and stitched coefficients, \(b_n - b_{n,s}\). However, the physically relevant case involves the resolvent in the limit \(\Im(z) \to 0\), since hydrodynamical features are determined by low-frequency behaviour. In this context, we formulated a criterion for uniform convergence along the imaginary axis and presented a formula linking the Lanczos coefficients to the spectral function's value at the origin, Eq.~\eqref{Conjecture}, consistent with the results in Ref. \cite{PhysRevB.110.104413}.

We have shown that slowly decaying staggered terms in the Lanczos coefficients’ asymptotics can slow the convergence of the approximation. Specifically, we have considered the case of approximating finite and nonzero Green's functions at the origin, when the recurrence coefficients follow the OGH in \(d > 1\), \(b_n = \alpha n + \gamma + o(1)\), employing the Meixner-Pollaczek recurrence coefficients for stitching. Our findings reveal two distinct regimes:
\begin{itemize}
  \item Relevant staggering: If the staggered term decays slowly, Meixner-Pollaczek stitching provides no advantage over constant stitching. Here, ``relevant" refers to cases where the staggered term follows a power-law decay in $n$ that is slower than $1/n$.
  \item Irrelevant staggering: If the staggered term decays rapidly, Meixner-Pollaczek stitching results in faster convergence than constant stitching. Here, ``irrelevant" refers to cases where the staggered term decays faster than $1/n$.
\end{itemize}
These staggered terms are linked to the non-analyticity of the spectral function. Specifically, for Lanczos coefficients obeying the OGH, we showed (non-rigorously) that if the spectral function has a finite number of derivatives at the origin, the staggered term decays as a power of \((\log n)^{-1}\). The higher the degree of differentiability of the spectral function at the origin, the faster the decay of the staggered term. In the cases of most physical interest (e.g., calculating a diffusion constant) we expect the error in the above stitching procedures to scale as $1/\mathrm{poly log}(N)$. This suggests that in order to achieve a target error of $\epsilon$ in the diffusion constant, $N\propto \exp(\mathrm{poly}(1/\epsilon))$ is required. 

There are a number of possible future directions. First of all, the motivation for considering staggered subleading corrections to the operator growth hypothesis is inspired by known Lanczos sequences, numerical results, and the results of a work connecting Lanczos coefficients to a Riemann-Hilbert problem \cite{lunt2025emergentrandommatrixuniversality}.
But a deeper understanding as to why subleading corrections to Lanczos coefficients takes this form would be welcome. Secondly, it would be worth further studying the validity of our fundamental Assumption~\ref{Convergence assumption}, which proposes a recipe for calculating Green's functions in the thermodynamic limit. Lastly, we provide some estimates for the Lanczos iterations $N$ required to achieve a given accuracy $\epsilon$ in a Green's function; but we have not investigated the computational complexity of calculating the first $N$ such coefficients. Naively the cost is $\exp(O(N))$, but \cite{PhysRevB.102.035147} suggest the cost might be significantly lower ($\mathrm{poly}(N)$); a definitive answer to that question would tell us the memory/CPU resources required by the Lanczos method to achieve a target accuracy $\epsilon$.

\section{Acknowledgements}
G.P. is supported by an EPSRC
studentship. C.K. is supported by a UKRI Future Leaders Fellowship through MR/T040947/2 and MR/Z000297/1. O.L.\ is supported by EPSRC through grant number EP/Y005058/2.
Some of the numerical computations were performed using King’s College London’s CREATE cluster \cite{create2025}.

%

\onecolumngrid
\newpage
\appendix

\section{Derivations of key results}

In this section, we present proofs of some results that are used in the main part of the paper. 
\subsection{Phase of the Cauchy transform}
The first result is that when $y>0$ the following holds:
\begin{equation}
    C_n(-\mathrm{i}y) = \mathrm{i}^{n+1}c_n(y) \hspace{5pt}
    \label{eqn: Cauchy phase}
\end{equation}
with $c_n(y)\geq 0$. Recall the definition of the Cauchy transform, $C_n(z)$:
\begin{equation}
C_n(z) = \langle O_n, (z-\mathcal{L})^{-1} O_0 \rangle = \int_{\mathbb{R}}\mathrm{d}\mu(x) \frac{p_n(x)}{z-x}   \hspace{5pt}.
\end{equation}
In order to prove the result, it is convenient to use the following identity
\begin{equation}
\int_{\mathbb{R}}\mathrm{d}\mu(x) \frac{p_n(z)-p_{n}(x)}{z-x}p_n(x) = 0 \hspace{5pt},    
\end{equation}
which follows from the fact that $p_n(x)$ is orthogonal to any polynomial in $x$ of degree less than $n$. This identity allows us to obtain an alternative expression for $C_n(z)$:
\begin{equation}
C_n(z) =    \frac{1}{p_n(z)}\int_{\mathbb{R}}\mathrm{d}\mu(x) \frac{p^2_n(x)}{z-x}  \hspace{5pt}.
\end{equation}
Substituting $z=-\mathrm{i}y$ yields
\begin{equation}
C_n(-\mathrm{i}y) =    \frac{\mathrm{i}y}{p_n(-\mathrm{i}y)}\int_{\mathbb{R}}\mathrm{d}\mu(x) \frac{p^2_n(x)}{y^2+x^2} \hspace{5pt}, 
\label{eqn: Cauchy transform}
\end{equation}
where we have used the fact that the spectral function is even. This expression has the advantage that the integral is always positive. It is easy to verify, using the three-term recurrence relation, that the orthogonal polynomials, for $y>0$, are of the form 
\begin{equation}
    p_n(-\mathrm{i}y) = \mathrm{i}^{-n} \Tilde{p}_{n}(y) \hspace{10pt} \Tilde{p}_n(y)>0 \hspace{5pt},
\end{equation}
which can then be inserted in Eq.~\eqref{eqn: Cauchy transform}. This leads to Eq.~\eqref{eqn: Cauchy phase} which concludes the proof.
\subsection{Limit of the infinite product for Meixner–Pollaczek}
We show that the following limit
\begin{equation}
\mathrm{i} \lim_{N \to \infty }\frac{1}{b_{N,s}}\prod_{k=1}^{\lfloor N/2\rfloor}\frac{b_{2k,s}^2}{b_{2k-1,s}^2} = G_s(-\mathrm{i}0^{+}) 
\label{eqn: MP infinite product}
\end{equation}
holds for the recurrence coefficients defining the Meixner–Pollaczek polynomials. 
First, we obtain an explicit expression for the right-hand side of Eq.~\eqref{eqn: MP infinite product}: $G_s(-\mathrm{i}0^{+})$. We use the spectral function of the Meixner–Pollaczek polynomials, which is given by \cite{koekoek2010hypergeometric}:
\begin{equation}
    \rho_s(x;\eta) = \frac{2^{\eta-2}}{\pi \alpha\Gamma(\eta)}\Bigl|\Gamma\Bigl(\frac{\eta + ix/\alpha}{2}\Bigl)\Bigl|^2 \hspace{5pt}.
\end{equation}
Setting $x=0$ and multiplying by $\mathrm{i}\pi$ yields:
\begin{equation}
G_{s}(-\mathrm{i}0^{+}) = \mathrm{i}\pi\rho_s(0;\eta) = \mathrm{i}\frac{2^{\eta-2} (\Gamma(\eta/2))^2}{\alpha \Gamma(\eta)}\hspace{5pt}.  
\label{eqn: MP zero-frequency Green's function}
\end{equation}
The task is now to reduce the left-hand side of Eq.~\eqref{eqn: MP infinite product} to Eq.~\eqref{eqn: MP zero-frequency Green's function}. Without loss of generality let $N= 2M$, where $M$ is integer, we then have
\begin{align}
\mathrm{i}\Pi_{2M,s} = & \;
\mathrm{i} \frac{1}{b_{2M,s}}\prod_{k=1}^{M}\frac{b_{2k,s}^2}{b_{2k-1,s}^2} = \mathrm{i} \frac{1}{\alpha \sqrt{2M(2M-1+\eta)}} \prod_{k=1}^{M} \frac{2k(2k-1+\eta)}{(2k-1)(2k-2+\eta)} \nonumber \\
= & \; \mathrm{i} \frac{1}{\alpha \sqrt{2M(2M-1+\eta)}} \frac{\sqrt{\pi } \Gamma (M+1) \Gamma(\frac{\eta+1}{2}+M)\Gamma(\eta/2)}{\Gamma(M+1/2) \Gamma(\eta/2+M) \Gamma(\frac{\eta+1}{2})} \hspace{5pt}.
\end{align}
The last equality has been obtained using the recurrence formula \cite{abramowitz1948handbook} for the gamma function which leads to:
\begin{equation}
\prod_{k=1}^{M}(2k-\alpha) = 2^M\frac{\Gamma(\frac{\alpha}{2}+M+1)}{\Gamma(\frac{\alpha}{2}+1)} \hspace{5pt}.
\end{equation}
We can then use the asymptotic expansion of the gamma function \cite{abramowitz1948handbook}
\begin{equation}
    \Gamma(M+a) \sim \left(\frac{M}{e}\right)^{M}M^{a-1/2}\sqrt{2\pi} \hspace{20pt} M \to \infty \hspace{5pt},
\end{equation}
to deduce that
\begin{equation}
    \mathrm{i}\Pi_{2M,s} \sim \mathrm{i}\frac{\Gamma(\eta/2)}{\Gamma(\frac{\eta+1}{2})}\frac{\sqrt{\pi}}{2\alpha} = \mathrm{i} \frac{2^{\eta-2}}{\alpha} \frac{(\Gamma(\eta/2))^2}{\Gamma(\eta)} \hspace{5pt},
    \label{eqn: Asymptotic of MP product}
\end{equation}
where we have used Legendre duplication formula in the second equality \cite{abramowitz1948handbook}. By comparing Eq.~\eqref{eqn: MP zero-frequency Green's function} and Eq.~\eqref{eqn: Asymptotic of MP product}, we can see that  $\mathrm{i}\Pi_{N,s} \to G_s(-\mathrm{i}0^{+})$ as claimed.
\subsection{Assumption~\ref{Convergence assumption} in terms of the \texorpdfstring{$n$}{n}-th level Green's function}
\label{Convergence of b_nG^n}
A consequence of Assumption~\ref{Convergence assumption} is the following limit:
\begin{equation}
    \lim_{N \to \infty}\mathrm{i}\Pi_N = G(-\mathrm{i}0^{+}) \hspace{5pt},
    \label{Appendix Conjecture}
\end{equation}
expressed in terms of the infinite product
\begin{equation}
    \Pi_m = \frac{1}{b_{m}}\prod_{j=1}^{\lfloor m/2\rfloor}\frac{b_{2j}^2}{b_{2j-1}^2} \hspace{5pt}.
    \label{eqn: Appendix Product}
\end{equation}
We show that this result, Eq.~\eqref{Appendix Conjecture}, is equivalent to:
\begin{equation} \lim_{n \to \infty} b_n G^{(n)}(-\mathrm{i}0^{+}) = \mathrm{i} \hspace{5pt}.
\label{eqn: zero-frequency limit of G_n}
\end{equation}
Recall that \( G^{(n)}(z) \) is the n-th level Green's function, defined as
\begin{equation}
G^{(n)}(z) = \langle O_n, (z-\mathcal{P}_{\geq n}\mathcal{L}\mathcal{P}_{\geq n})^{-1} O_n \rangle,
\end{equation}
and \( \mathcal{P}_{\geq n} \) is a projector into the subspace spanned by the basis vectors \( \{O_j\}_{j=n}^{\infty} \).

To proceed, we need the following relation:
\begin{equation}
b_n G^{(n)}(z) = \frac{C_n(z)}{C_{n-1}(z)} \hspace{5pt}.
\label{eqn: Ratio of Cauchy}
\end{equation}
This result is proved in Ref.~\cite{VANASSCHE1991237} (Eq. (3.7)). However, we present a different proof for the reader's convenience. Our technique involves verifying that both the ratio of successive Cauchy transforms, $C_n(z)/C_{n-1}(z)$, and $b_n G^{(n)}(z)$ satisfy the same recurrence relation with the same initial conditions, and hence they are equal to each other. As mentioned in the paper, the recurrence relations are
\begin{equation}
    G^{(n)}(z) = \frac{1}{z-b_{n+1}^2G^{(n+1)}(z)} \hspace{5pt}, 
    \label{eqn: Recurrence Gn}
\end{equation}
with initial conditions $G^{(0)}(z) = G(z)$ and
\begin{equation}
    b_{n+1}C_{n+1}(z) = z C_n(z) -b_nC_{n-1}(z) \hspace{5pt},
\end{equation}
with initial conditions $C_0(z)=G(z)$ and $C_1(z) = (zG(z)-1)/b_1$. Hence, the ratio $r_n(z) =  C_n(z)/C_{n-1}(z)$ satisfies
\begin{equation}
    r_{n+1}(z) = \frac{z}{b_{n+1}}-\frac{b_{n}}{b_{n+1}}\frac{1}{r_n(z)} \implies \frac{r_n(z)}{b_n} = \frac{1}{z-b^2_{n+1}\frac{r_{n+1}(z)}{b_{n+1}}}
    \label{eqn: Recurrence ratio}
\end{equation}
with $r_0(z) = C_0(z) = G(z)$. Comparing Eq.~\eqref{eqn: Recurrence ratio} and Eq.~\eqref{eqn: Recurrence Gn}, we conclude that $r_n(z)/b_n = G^{(n)}(z)$ which proves Eq.~\eqref{eqn: Ratio of Cauchy}. Having obtained this relation, we can now justify Eq.~\eqref{eqn: zero-frequency limit of G_n}. Setting $z \to -\mathrm{i}0^{+}$ in the recurrence relation for $r_n(z)$, Eq.~\eqref{eqn: Recurrence ratio}, and iterating yields
\begin{equation}
    r_n(-\mathrm{i}0^{+}) = \begin{cases}
        -\frac{\Pi_n}{G(-\mathrm{i}0^{+})}& \text{n odd} \\
         \frac{G(-\mathrm{i}0^{+})}{\Pi_n}& \text{n even}
    \end{cases} \hspace{5pt}.
\end{equation}
According to Eq.~\eqref{Conjecture},which in turn follows from Assumption~\ref{Convergence assumption}, we conclude that $r_n(-\mathrm{i}0^{+}) = b_nG^{(n)}(-\mathrm{i}0^{+}) \to \mathrm{i}$.

We also argue that a similar result holds at all frequencies $z$ such that $\Im(z) \neq 0$:
\begin{equation}
    \lim_{n \to \infty} b_n G^{(n)}(z) = -\mathrm{sgn}(\Im(z))\mathrm{i} \hspace{5pt}.
    \label{eqn: Limit valid for all frequencies}
\end{equation}

In this case, we use the recurrence relation satisfied by $r_n(z)$, Eq.~\eqref{eqn: Recurrence ratio}, which we manipulate to obtain:
\begin{equation}
 r_{n+1}(z)r_n(z) = \frac{z r_n(z)}{b_{n+1}} - \frac{b_n}{b_{n+1}} \hspace{5pt}.
 \label{eqn: Ratio manipulation}
\end{equation}
We assume that $b_n \to \infty$ and $\frac{b_n}{b_{n+1}} \to 1$, which hold under the OGH. Furthermore, we assume that $r_n(z)$ remains bounded as a function of $n$, so that the first term on the right-hand side of Eq.~\eqref{eqn: Ratio manipulation} vanishes in the limit of large $n$.
Hence, this analysis yields the following limit:
\begin{equation}
    \lim_{n \to \infty} r_{n+1}(z) r_n(z) = -1 \hspace{5pt}.
    \label{eqn: Limit}
\end{equation}
Assuming that the limit $\lim_{n\to \infty}r_n(z)$ exists, we conclude that
\begin{equation}
    \lim_{n\to \infty}r_n(z) = \lim_{n\to \infty}b_nG^{(n)}(z) = \pm i \hspace{5pt},
\end{equation}
which is consistent with the earlier assumption that $r_n(z)$ is bounded in $n$.
The sign depends on the imaginary part of $z$, as the imaginary part of the Green's function has a definite sign: $\mathrm{sgn}(\Im(G(z))) = -\mathrm{sgn}(\Im(z))$. This property allows us to fix the sign, thereby proving Eq.~\eqref{eqn: Limit valid for all frequencies} for $\Im(z) \neq 0$.

In conclusion, we have shown that Assumption~\ref{Convergence assumption} is equivalent to Eq.~\eqref{eqn: zero-frequency limit of G_n}. 
Furthermore, we have argued that a similar limit holds for all frequencies $z$ such that $\Im(z) \neq 0$: Eq.~\eqref{eqn: Limit valid for all frequencies}. This equation holds, provided that the limit exists and certain additional assumptions regarding the Lanczos coefficients are satisfied, as is the case under the OGH.
\subsection{Convergence of the infinite product and asymptotic expansion}
In this section we obtain a criterion for the convergence of the product in Eq.~\eqref{eqn: Appendix Product}.
The standard technique is to convert the product into a summation and then approximate sums with integrals, which is justified by the Euler–Maclaurin formula. We set without loss of generality $m = 2n+1$ and rewrite $\Pi_{2n+1}$ as
\begin{equation}
    \Pi_{2n+1} = \exp \Biggl(\sum_{j=1}^{n}2\log\left(1+\frac{b_{2j}-b_{2j-1}}{b_{2j-1}}\right) - \log b_{2n+1}\Biggl) \hspace{5pt}. \label{eqn: Sum}
\end{equation}

We proceed by expanding the logarithm, assuming that $\frac{b_{2n}-b_{2n-1}}{b_{2n-1}} \to 0$, which is true for Lanczos coefficient satisfying the OGH. Some extra care is needed if the asymptotic expansion of $b_n$ contains staggered terms. Let $b_n = f_n + (-1)^ns_n$ with $s_n/f_n \to 0$; then we get
\begin{eqnarray}
    \Pi_{2n+1} &\propto& \exp \Biggl(\sum_{j=\mathcal{O}(1)}^{n}2\log \left(1+\frac{f_{2j}-f_{2j-1}}{f_{2j-1}} + \frac{s_{2j}+s_{2j-1}}{f_{2j-1}}\right) - \log f_{2n+1}\Biggl) \nonumber\\
    &\propto& \exp \Biggl(\sum_{j=\mathcal{O}(1)}^{n}2\left(\frac{f'_{2j}}{f_{2j}} + \frac{s_{2j}+s_{2j-1}}{f_{2j-1}}\right) - \log f_{2n+1}\Biggl) \propto \exp \Biggl(2\int_{\mathcal{O}(1)}^{n} \mathrm{d}x \left(\frac{f'_{2x}}{f_{2x}} + \frac{s_{2x}+s_{2x-1}}{f_{2x-1}}\right) - \log f_{2n+1}\Biggl) \nonumber\\
    &\propto& \exp \Biggl(2\int_{\mathcal{O}(1)}^{n} \mathrm{d}x  \frac{s_{2x}+s_{2x-1}}{f_{2x-1}}\Biggl) \propto \exp \Biggl(4\int_{\mathcal{O}(1)}^{n} \mathrm{d}x \frac{s_{2x}}{f_{2x}}\Biggl)  \hspace{5pt},
    \label{eqn: Product Asymptotic}
\end{eqnarray}
where we have assumed that $f_{n}$ and $s_n$ do not grow too fast so that $s_{2n}+s_{2n-1} \sim 2s_{2n}$. Furthermore, we have also assumed that the finite difference $f_{2j}-f_{2j-1}$ can be approximated by a derivative.
This simple calculation shows that staggering might cause divergences in the spectral function at the origin. In particular, if the integral $\int_{\mathcal{O}(1)}^{n} \mathrm{d}x s_{2x}/f_{2x} \to \pm \infty$ diverges then $\Pi_{2n+1} \to \infty$ or $\Pi_{2n+1} \to 0$. On the other hand, if there is no staggering, then the product converges and the subleading term is generally $\mathcal{O}(n^{-1})$. This correction arises, for example, from further corrections to the Euler-Maclaurin formula.
\subsection{Upper bound on the product of Cauchy transforms}
In this discussion, we aim to demonstrate that the following inequality holds:  
\begin{equation}  
c_{n}(y)c_{n\pm 1}(y;N) \leq \frac{M}{b_{n}}, 
\label{eqn: Bound on Cauchy product}
\end{equation}  
under the assumptions outlined in the main part of the paper. Specifically, we assume that the spectral functions $\rho(x)$ and $\rho_s(x)$, which correspond to the sequences $\{b_n\}$ and $\{b_{n,s}\}$, are finite and non-zero at the origin, and that the ratio $b_N/b_{N,s} \to 1$ as $N \to \infty$.
Recall that $N$ is the level at which the stitching approximation is performed, $c_n(y)$ is related to the Cauchy transform associated with the sequence $\{b_n\}$ and $c_n(y;N)$ to $\{b_n\}_{n=1}^N \cup \{b_{n,s}\}_{n=N+1}^{\infty}$.

We start with the following general result, proved in the main part of the paper,
\begin{eqnarray}
c_{2m+1}(y) &\leq&   c_1(y) \prod_{k=1}^{m
}\frac{b_{2k}}{b_{2k+1}} \nonumber\\
c_{2m}(y) &\leq&   c_0(y) \prod_{k=1}^{m}\frac{b_{2k-1}}{b_{2k}} \hspace{5pt},
\end{eqnarray}
which holds for any sequence $\{b_n\}$. It is useful to define
\begin{equation}
    b_{j,s}(N) = 
    \begin{cases}
        b_{j} & j\leq N \\
        b_{j,s} & j>N
    \end{cases} \hspace{5pt}.
\end{equation}
We begin by considering the product $c_{2m}(y)c_{2m+1}(y;N) $ with $m>(N-1)/2$ and we assume that $N$ is odd without loss of generality:
\begin{eqnarray}
c_{2m}(y)c_{2m+1}(y;N) &\leq& c_{0}(y)c_{1}(y;N)\Bigl[\prod_{k=1}^{m}\frac{b_{2k-1}}{b_{2k}}\Bigl]  \Biggl[\prod_{k=1}^{m}\frac{b_{2k,s}(N)}{b_{2k+1,s}(N)}\Biggl] \nonumber\\
&\leq&  C \Biggl[\prod_{k=1}^{m}\frac{b_{2k-1}}{b_{2k}}\Biggl]\Biggl[\prod_{k=1}^{(N-1)/2}\frac{b_{2k}}{b_{2k+1}} \Biggl]\Biggl[\prod_{k=(N+1)/2 }^{m}\frac{b_{2k,s}}{b_{2k+1,s}} \Biggl] \nonumber\\
&=& C \Biggl[\prod_{k=1}^{m}\frac{b_{2k-1}}{b_{2k}} \Biggl]\Biggl[\prod_{k=1}^{m}\frac{b_{2k,s}}{b_{2k+1,s}} \Biggl] \Biggl[\prod_{k=1}^{(N-1)/2}\frac{b_{2k}}{b_{2k+1}}\Biggl]
 \Biggl[\prod_{k=1}^{(N-1)/2}\frac{b_{2k+1,s}}{b_{2k,s}} \Biggl] \hspace{5pt}.
\label{eqn: stitching inequality}
\end{eqnarray}
In the second line, we used the fact that if the stitching approximation converges then $\lim_{n \to \infty}c_0(y;n) = c_{0}(y)$ and $c_0(y)$ is bounded by assumption. This follows because we focus on Green's functions, $G(z)$, that are finite and non-zero at the origin; furthermore, $G(z) \leq 1/|\Im(z)|$ which is enough to conclude that $C_0(-y) = G(-\mathrm{i}y)$ is bounded in $y$. Consequently, $c_0(y;N)$ and $c_1(y;N)$ are both bounded by some constants that are independent of $y$ and $N$, whose product is an unknown constant $C$. The remaining manipulations decoupled the $m$ and $N$ dependence as shown in the third line. The task is to check that the third line does not grow unbounded when $m$ and $N$ get arbitrarily large. This can be done using the following limits
\begin{equation}
\lim_{m \to \infty}\frac{1}{b_{2m+1}}\prod_{j=1}^{m}\frac{b_{2j}^2}{b_{2j-1}^2} = \pi \rho(0)  \hspace{10pt} \lim_{m \to \infty}\frac{1}{b_{2m+1,s}}\prod_{j=1}^{m}\frac{b_{2j,s}^2}{b_{2j-1,s}^2}  = \pi \rho_s(0)
\end{equation}
where $\rho(0)$ and $\rho_s(0)$ are the corresponding spectral functions, which are finite and non-zero. The first limit follows by Assumption~\ref{Convergence assumption} in the main text, whereas the second limit can be proved explicitly (see Eq. \eqref{eqn: MP infinite product}).
To begin, let us consider the large $N$ limit of Eq.~\eqref{eqn: stitching inequality}. Using the previous limits we obtain
\begin{equation}
\prod_{k=1}^{(N-1)/2} \frac{b_{2k}}{b_{2k+1}}  \sim  b_{1}\sqrt{\frac{\pi \rho(0)}{b_{N}}}  \hspace{20pt} \prod_{k=1}^{(N-1)/2} \frac{b_{2k+1,s}}{b_{2k,s}}  \sim  \frac{1}{b_{1,s}}\sqrt{\frac{b_{N,s}}{\pi \rho_s(0)}} \hspace{5pt}.
\end{equation}
The product of these two terms is bounded since $b_{N}/b_{N,s} \to 1$ holds by assumption. 
On the other hand,  when $m$ is large 
\begin{equation}
\prod_{k=1}^{m}\frac{b_{2k-1}}{b_{2k}} \sim \frac{1}{\sqrt{\pi \rho(0) b_{2m}}} \hspace{20pt} \prod_{k=1}^{m}\frac{b_{2k,s}}{b_{2k+1,s}} \sim b_{1,s}\sqrt{\frac{\pi \rho_s(0)}{b_{2m,s}}} \hspace{5pt}.
\end{equation}
Therefore, we conclude that, when $m \to \infty$, $c_{2m}(0)c_{2m+1}(0;N) = \mathcal{O}(1/b_{2m})$; hence, the $N$ dependence disappears in the upper bound.

Similarly, we can repeat the procedure to find an upper bound for $c_{2m}(y;N)c_{2m+1}(y)$, $c_{2m}(y)c_{2m-1}(y;N)$, and $c_{2m}(y;N)c_{2m-1}(y)$, all of which yield the same upper bound $\mathcal{O}(b_{2m}^{-1})$. Consequently, we obtain the upper bound in Eq.~\eqref{eqn: Bound on Cauchy product}. This result is important because it allows us to apply the Weierstrass M-test to establish a sufficient condition for the uniform convergence of the stitching approximation on the imaginary axis. The uniform convergence, in turn, simplifies the analysis of the rate of convergence of the stitching approximation.

\section{Truncation of continued fractions} 
In this section, we justify the claim in the main text that truncating the continued fraction is not a good approximation in the low-frequency regime. Specifically, we show that the rate of convergence worsens as the real axis is approached.

Mathematically, truncating the continued fraction at level $N$ consists of setting $b_{N}=0$. We can obtain a compact expression by iterating the M\"{o}bius transformation: 
\begin{equation}
    G(z) = \Omega_{n}\Bigl(G^{(n)}(z)\Bigl) \hspace{5pt}.
\end{equation}
The M\"{o}bius transformation $\Omega_n$ can be represented as a $2\times 2$ matrix
\begin{equation}
    \Omega_{n} = 
\begin{bmatrix}
    -b_nq_{n-1}(z)       & q_n(z) \\
     -b_np_{n-1}(z)      & p_n(z)
\end{bmatrix}
\end{equation}
where a representation of a generic element of the M\"{o}bius group $z \to \frac{az+b}{cz+d}$ is given by the matrix $\begin{bmatrix}
    a       & b \\
    c      & d
\end{bmatrix}$.
The polynomials $p_n$ and $q_n$ satisfy the same three-term recurrence relation with different initial conditions $\{p_{-1}(z)=0, p_{0}(z) = 1\}$ and $\{q_{0}(z)=0, q_{1}(z) = 1/b_{1}\}$. The polynomial $q_n(z)$ is of degree $n-1$ and is known as secondary polynomial \cite{simon2015operator}. These polynomials, after rescaling $\Tilde{q}_{n-1}(z) = q_n(z)b_1$,  constitute a new set of orthogonal polynomials with respect to the new spectral function $d\Tilde{\mu}(x) = d\mu(x)/(|G(x)|^2b_1^2)$; they are associated with the tridiagonal matrix with coefficients $\{b_{j}\}_{j=2}^{\infty}$. Furthermore, the polynomials $\{p_n(z),q_n(z)\}$ can be used to write a solution of any recurrence relations of the form $b_{n+1}h_{n+1}(z) = zh_n(z)-b_{n}h_{n-1}(z)$, with arbitrary initial conditions $h_{-1}(z) = f_{-1}(z)$, $h_{0}(z) = f_{0}(z)$: $h_n(z) = p_n(z)f_{0}(z)-q_n(z)f_{-1}(z)$.

Truncating the continued fraction at level $N$ is an approximation that requires setting $b_{N}=0$ or equivalently $G^{(N)}(z) = 0$:
\begin{equation}
    G_{N,t}(z) = \frac{q_{N}{(z)}}{p_{N}(z)} \hspace{5pt}.
\end{equation}
Setting $z=0$ in the above expression gives either infinity or zero depending on the parity of $N$. In view of this fact, the limit $\lim_{N \to \infty}G_{N,t}(0)$ does not exist. Therefore, truncation, unlike the stitching approximation, does not converge uniformly in $z$. However, it is still a useful approximation to the Green's function when $\Im(z) \neq 0$ especially when $\Im(z)$ is sufficiently large. In fact, we will show that the rate of convergence improves as $\Im(z)$ increases.

In order to find the rate of convergence as $N \to \infty$, we use the following representation for the secondary polynomials \cite{simon2015operator}
\begin{equation}
    q_n(z) = \int_{\mathbb{R}} \mathrm{d}\mu(x) \frac{p_{n}(z)-p_n(x)}{z-x}
\end{equation}
which implies that the error in truncating the continued fraction at level $n$ is given by
\begin{equation}
    e_{N,t}(z) = \frac{C_N(z)}{p_N(z)} \hspace{5pt}.
    \label{eqn: Truncation error level n}
\end{equation}
Therefore, the problem has been reduced to finding the asymptotic forms of $p_n(z)$ and its Cauchy transform $C_n(z)$. In general, this problem is hard to solve; the most common method used to tackle it is the Riemann-Hilbert method introduce first by Fokas, Its, and Kitaev \cite{fokas1992isomonodromy}. However, for the sake of simplicity, we will employ a less rigorous method. We consider the generic recurrence relation for the ratio $r_n(z)=h_n(z)/h_{n-1}(z)$ where $h_n(z)$ can be either $p_n(z)$ or $C_n(z)$ since they obey the same recursion relation
\begin{equation}
    r_{n+1}(z) = \frac{z}{b_{n+1}}-\frac{b_{n}}{b_{n+1}}\frac{1}{r_n(z)} \hspace{5pt},
\end{equation}
but with different initial conditions.
Recall that, if we assume that $b_n$ satisfies $b_n \to \infty$ and $b_n/b_{n+1} \to 1$, which is true if the OGH holds, and that $r_{n}(z)$ is bounded as a function of $n$ we obtain
\begin{equation}
    \lim_{n \to \infty} r_{n+1}(z)r_{n}(z) = -1 \hspace{5pt}.
    \label{eqn: Limit_reminder}
\end{equation}
Under the assumption that $\lim_{n \to \infty} r_{n}(z)$ exists, for $z$ in some region of the complex plane, we conclude that
\begin{equation}
\lim_{n \to \infty} r_{n}(z)  = \pm \mathrm{i} \hspace{5pt}, 
\label{eqn: Existence of limit}
\end{equation}
where the sign depends on the sign of $\Im(z)$ and whether $r_n(z)$ is a ratio of two successive polynomials or two successive Cauchy transforms. The aim is now to find the correction to the limit in Eq.~\eqref{eqn: Limit_reminder}, which will then allows us to extract the asymptotic of $p_n(z)$ and $C_n(z)$. We focus on the case where $b_n = \alpha n + \mathcal{O}(1)$ which corresponds to the OGH in $d>1$. 
Let
\begin{equation}
     r_{n+1}(z)r_{n}(z) = -1+\Delta_{n+1}(z) \hspace{20pt} \Delta_{n+1}(z) = o(1) \hspace{5pt},
\end{equation}
we will prove that $\Delta_{n} = \mathcal{O}(n^{-1})$. Plugging back into the recurrence relation yields
\begin{equation}
    -1+\Delta_{n+1}(z) = \frac{z}{b_{n+1}}r_n(z)-\frac{b_n}{b_{n+1}} \hspace{5pt}, 
\end{equation}
thus
\begin{equation}
    \lim_{n \to \infty}\frac{\Delta_{n+1}(z)}{n^{-1}} = \lim_{n \to \infty}\frac{z}{b_{n+1}n^{-1}}r_n(z)-\lim_{n \to \infty}\Bigl(\frac{b_n}{b_{n+1}}-1\Bigl)\frac{1}{n^{-1}} = \frac{\pm i z}{\alpha}+1 \hspace{5pt}.
\end{equation}
Therefore,
\begin{equation}
r_{n+1}(z)r_{n}(z)= \frac{h_{n+1}(z)}{h_{n-1}(z)}= -1+ \frac{1}{n}\frac{\pm \mathrm{i}z+\alpha}{\alpha}  +o(n^{-1})  \hspace{5pt}.
\label{eqn: Asymptotic recurrence ratio}
\end{equation}
Iterating the recurrence for $h_n(z)$ yields
\begin{equation}
    h_n(z) = A(z)\mathrm{i}^nn^{\frac{-\mathrm{i}z - \alpha}{2\alpha}}(1+o(1)) + B(z) (-\mathrm{i})^nn^{\frac{\mathrm{i}z - \alpha}{2\alpha}}(1+o(1)) \hspace{5pt}.
    \label{eqn: Asymptotic from recurrence}
\end{equation}   
where $A(z)$ and $B(z)$ are unknown functions that reflect the two possible limits in Eq.~\eqref{eqn: Existence of limit}. 
 
We first consider $C_n(z)$. Since this is bounded above by $|C_n(z)| \leq \frac{1}{|\Im(z)|}$, we conclude that $\lim_{n \to \infty}C_n(z)$ cannot diverge. Hence,
\begin{equation}
    C_n(z) = \begin{cases}
        C(z)\mathrm{i}^n n^{\frac{-\mathrm{i}z - \alpha}{2\alpha}}(1+o(1)) & \Im(z)<0 \\

        \bar{C}(z)\mathrm{i}^{-n} n^{\frac{\mathrm{i}z - \alpha}{2\alpha}}(1+o(1)) & \Im(z)>0 
    \end{cases} \hspace{5pt}, 
\end{equation}
where $C(z)$ is some unknown function. 

Similarly, using the fact that $p_n(z) \to \infty$ as $\Im(z) \to \infty$ we obtain
\begin{equation}
    p_n(z) = \begin{cases}
        p(z)\mathrm{i}^n n^{\frac{\mathrm{i}z - \alpha}{2\alpha}}(1+o(1)) & \Im(z)<0 \\

        \bar{p}(z)\mathrm{i}^{-n} n^{\frac{-\mathrm{i}z - \alpha}{2\alpha}}(1+o(1)) & \Im(z)>0 
    \end{cases} \hspace{5pt}.
    \label{eqn: Asymptotics polynomials}
\end{equation}
Therefore, we get, from Eq.~\eqref{eqn: Truncation error level n}, that the error in truncating a continued fraction, at level $N$, scales as
\begin{equation}
    |e_{N,t}(z)| = f(z) N^{-\frac{|\Im(z)|}{\alpha}}(1+o(1)) \hspace{5pt},
    \label{eqn: Truncation error}
\end{equation}
for some function $f(z)>0$. This formula shows that the error converges very slowly as $\Im(z) \to 0$ and the convergence improves as $\Im(z)$ increases. The fact that the convergence  worsens in the zero-frequency limit signals that truncation is not a good approximation to probe hydrodynamics, so it is not suitable to extract a diffusion constant, which motivates using the stitching approximation analysed in the main text.

\begin{figure}
\subfloat{%
  \includegraphics[width=0.5\columnwidth]{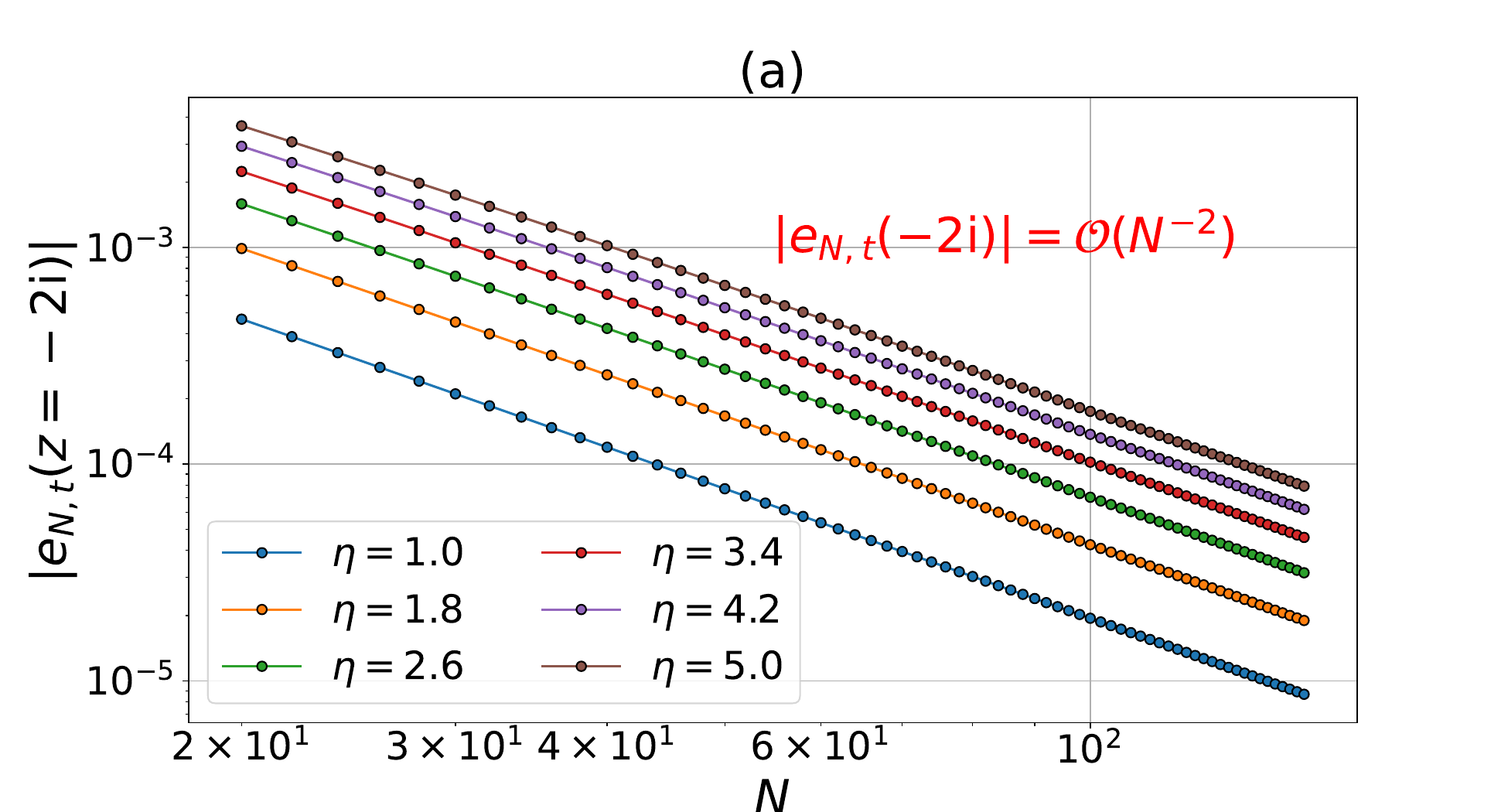}%
}\hspace*{\fill}%
\subfloat{%
  \includegraphics[width=0.5\columnwidth]{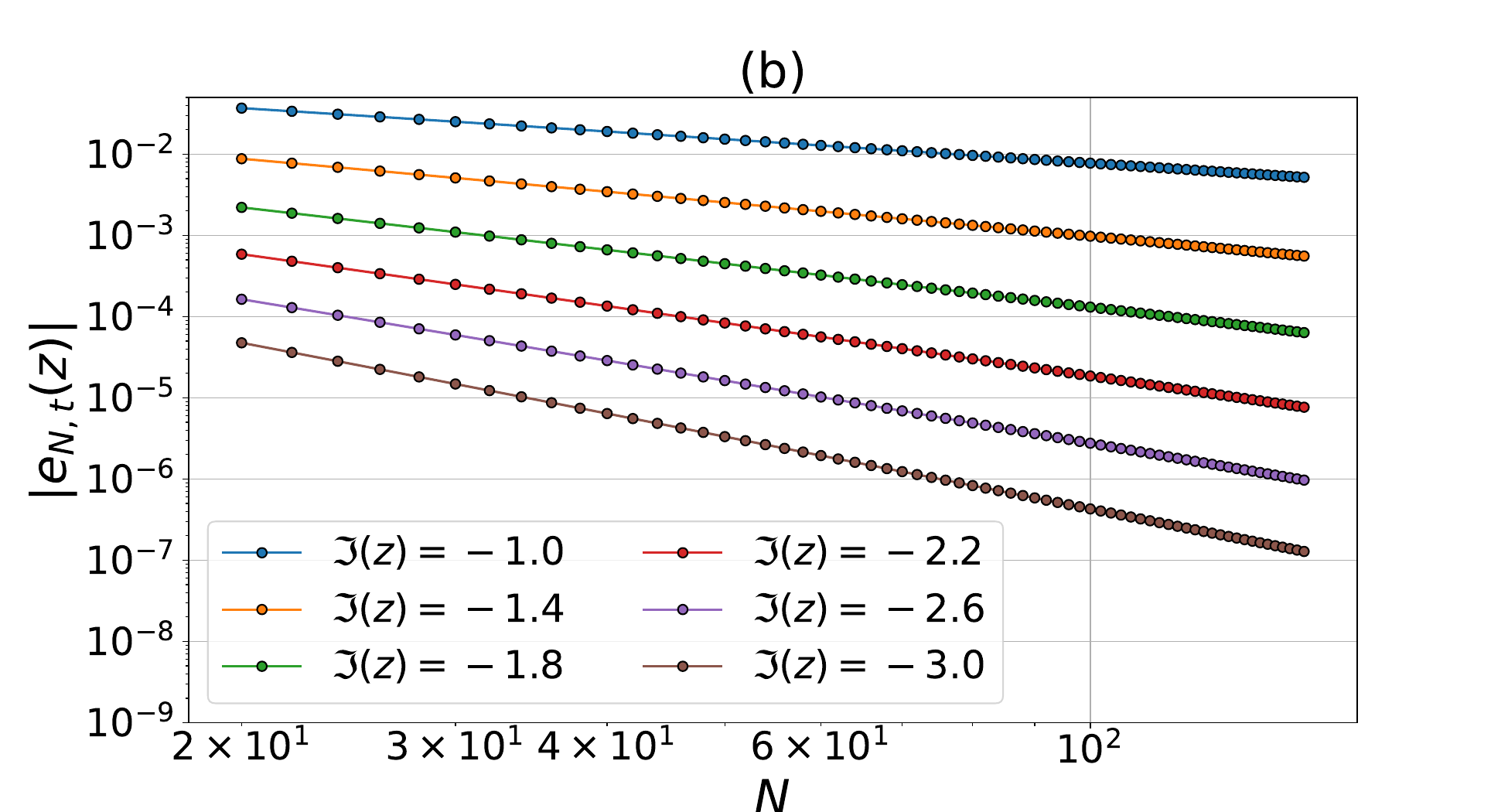}%
}

\medskip
\centering
\includegraphics[width=0.6\columnwidth]{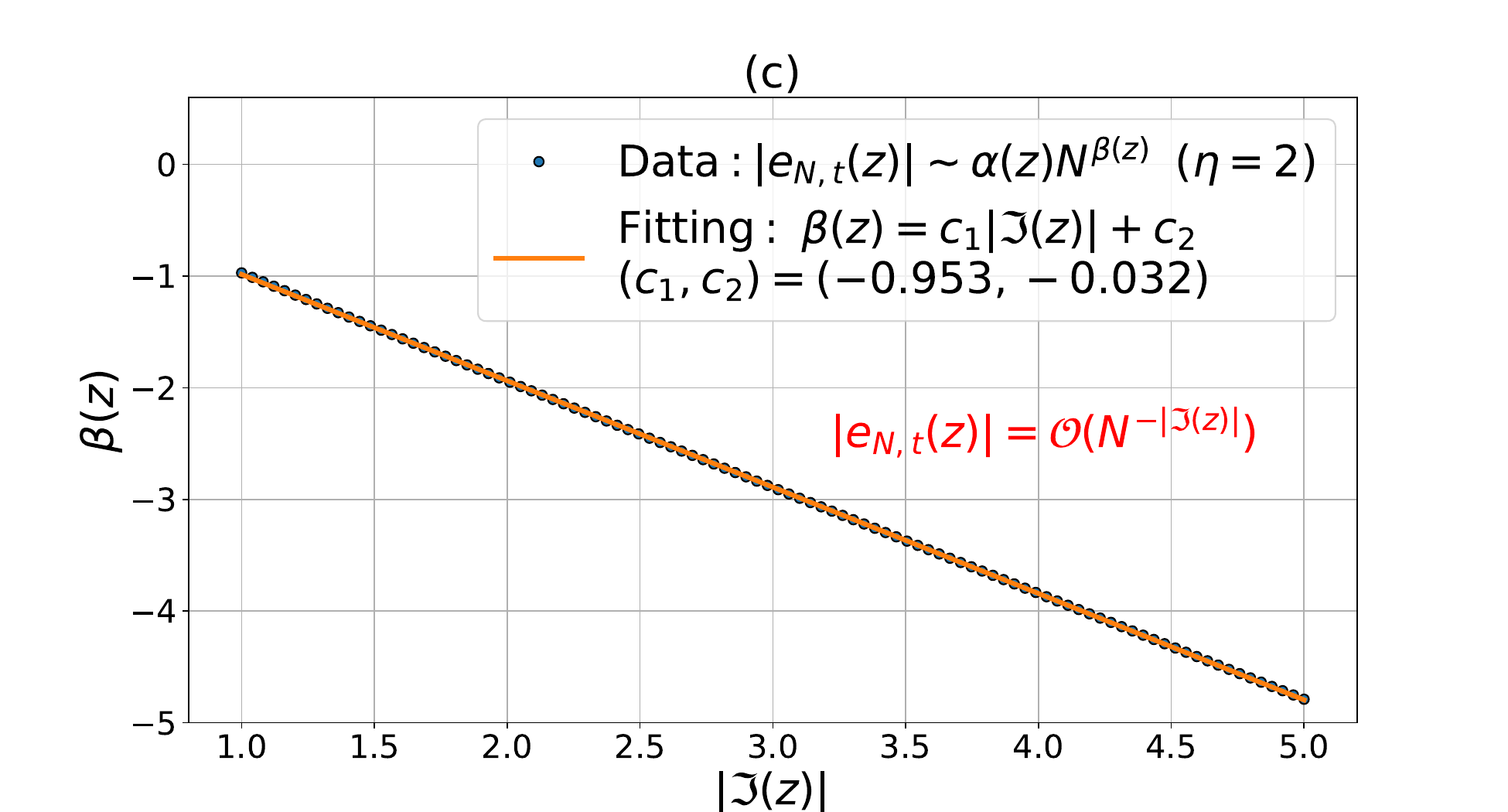}
    
\caption{Truncation error as a function of the truncation level $N$. (a) The error for some values of $\eta$, which is a parameter in the recurrence coefficients of the Meixner–Pollaczek polynomials, at a fixed $z=-2\mathrm{i}$. The particular value of $\eta$ does not affect the convergence rate. (b) Plot of the error as a function of $N$ for different values of $\Im(z)|$, for fixed $\eta=2$. The larger the value of $|\Im(z)|$ the faster the approximation converges, independently of the value of $\eta$. (c) Fitting to determine the exponent $\beta(z)$ defined as $e_{N,t}(z) \sim \alpha(z) N^{\beta(z)}$. The analytical prediction from Eq.~\eqref{eqn: Truncation error} is $\beta(z) = -|\Im(z)|$. Therefore, we expect the fitted line to have a slope of $-1$ and a zero intercept. The fitted parameters $c_1$ and $c_2$ are indeed close to these values.} 
\label{Truncation}
\end{figure}

In a similar fashion, we can calculate the asymptotic of the truncation error when $b_n \sim \alpha n/\log(n)$ (OGH in $d=1$) and the spectral function is finite at the origin. In this case, we find that
\begin{equation}
r_{n+1}(z)r_{n}(z)= \frac{h_{n+1}(z)}{h_{n-1}(z)}= -1+ \frac{1}{n}\Bigl[\frac{\pm iz \log n}{\alpha} + 1 - \frac{1}{\log n} \Bigl]  +O(n^{-2})  \hspace{5pt}.
\end{equation}
Hence we conclude that, for $d=1$, we have the following asymptotic expansions
\begin{equation}
    C_n(z) = \begin{cases}
        C(z)\mathrm{i}^n \sqrt{\frac{\log n}{n}} n^{-\frac{iz \log n}{4\alpha}}(1+o(1)) & \Im(z)<0 \\

        \bar{C}(z)\mathrm{i}^{-n} \sqrt{\frac{\log n}{n}} n^{\frac{\mathrm{i}z \log n}{4\alpha}}(1+o(1)) & \Im(z)>0 
    \end{cases} \hspace{20pt}
       p_n(z) = \begin{cases}
        p(z)\mathrm{i}^\mathrm{i} \sqrt{\frac{\log n}{n}} n^{\frac{\mathrm{i}z \log n}{4\alpha}}(1+o(1)) & \Im(z)<0 \\

        \bar{p}(z)\mathrm{i}^{-n} \sqrt{\frac{\log n}{n}} n^{-\frac{\mathrm{i}z \log n}{4\alpha}}(1+o(1)) & \Im(z)>0 
    \end{cases} \hspace{5pt},
\end{equation}
\begin{equation}
    |e_{N,t}(z)| = f(z) N^{-\frac{|\Im(z)|}{2\alpha}\log N}(1+o(1)) \hspace{5pt}.
\end{equation}
This converges slightly faster than $b_n \sim \alpha n$ but still the convergence worsens as $|\Im(z)| \to 0$. It is now clear that the limits $|\Im(z)| \to 0$ and $N \to \infty$ do not commute which implies that the convergence is not uniform in $z$. 
Numerical evidence for the solvable case of the Meixner–Pollaczek polynomials, $b_{n} = \sqrt{n(n-1 + \eta)}$, is presented in Fig. \ref{Truncation}. The results agree with the prediction of Eq.~\eqref{eqn: Truncation error}.

\section{Smoothness of the spectral function}
In this section, we determine what conditions should be satisfied by the Lanczos coefficients so that the $k$-th derivative is the first divergent derivative of the spectral function at the origin. This section is structured as follows. In the first part, we outline the strategy, which is based on using the consequence of Assumption~\ref{Convergence assumption} to obtain an approximation for the Green's function expressed in terms of the orthogonal polynomials. The aim is to study the asymptotic behaviour of the $k$-th derivative of such expression.

Before doing so, we need to examine the asymptotic behaviour of the derivatives of the polynomials at the origin, along with their corrections. This is the focus of the second part. Building on these results, the third part presents the main result of this section: a criterion on the staggered subleading correction to the Lanczos coefficients so that the $k$-th derivative is the first to diverge. The fourth part presents a supplementary calculation that illustrates why it is necessary to include the subleading correction of the derivatives of the polynomials examined in the second part. Finally, the last part presents some numerical results to support the analytical predictions presented in the third part.
\subsection{Strategy: approximating the M\"{o}bius transformation}
Recall that the spectral function, $\rho(x)$, and the Green's function, $G(z)$, are related via $\rho(x) = \pi^{-1}\lim_{\epsilon \to 0^{+}}\Im(G(x-\mathrm{i} \epsilon))$. Consequently, analysing the divergence of $\rho^{(k)}(0)$ is equivalent to analysing $G^{(k)}(-\mathrm{i}0^{+})$. Taking this into consideration, we propose a non-rigorous way to analyse this problem. First, we use the assumption that $b_nG^{(n)}(z) \to i$, which is valid when $\Im(z)<0$, and it is a consequence of Assumption~\ref{Convergence assumption} (see Subsection~\ref{Convergence of b_nG^n}). Consequently, we approximate $G(z) = \Omega_{2n}\Bigl(G^{(2n)}(z)\Bigl)$ with $G(z;2n) \equiv \Omega_{2n}(\mathrm{i}/b_{2n})$ which, after iterating the M\"{o}bius transformation, has the following expression:
\begin{align}
   G(z;2n) \equiv \Omega_{2n}(\mathrm{i}/{2n}) =  \frac{q_{2n}(z)-\mathrm{i}q_{2n-1}(z)}{p_{2n}(z)-\mathrm{i}p_{2n-1}(z)}
    = \frac{q_{2n}(z)p_{2n}(z)+q_{2n-1}(z)p_{2n-1}(z)}{p^2_{2n}(z)+p^2_{2n-1}(z)}+\mathrm{i}\frac{1}{b_{2n}\Bigl(p^2_{2n}(z)+p^2_{2n-1}(z)\Bigl)} \hspace{5pt},
    \label{eqn: G(z;n)}
\end{align}
where, in the last equality, we used the fact that $b_{2n+1}(q_{2n+1}(z)p_{2n}(z)-q_{2n}(z)p_{2n+1}(z)) = 1$ \cite{simon2015operator}. We have also used $2n$ so that the parity of the degree of the polynomial is easier to track.

We then compute the $k$-th derivative of $G(z;2n)$ at the origin and seek a condition for this to diverge when $n$ is large, which will constrain the subleading correction of $b_n$ in some way.
Notice, from Eq.~\eqref{eqn: G(z;n)}, that the real part of $G(z;2n)$ is an odd function, whereas the imaginary part is an  even function. This follows since $q_{j}(z)$ and $p_{j}(z)$ have definite parity because the spectral function is even. Specifically, $p_j(z)$ is an even (odd) function when $j$ is even (odd). On the other hand, $q_j(z)$ is an even (odd) function when $j$ is odd (even).
Hence, all even derivatives of $G(z;2n)$ at the origin are purely imaginary:
\begin{equation}
G^{(2l)}(0;2n)= \frac{\mathrm{i}}{b_{2n}} \partial^{2l}_z \frac{1}{p^2_{2n}(z)+p^2_{2n-1}(z)}\Biggl|_{z=0} \hspace{5pt}.   
\end{equation}
For simplicity, we consider only the even case; thus, we set the order of the first divergent derivative, $k$, equal to an even integer: $k = 2l$. The odd case is similar and leads to the same conclusions. Using Faà di Bruno's formula \cite{faa1855sullo}, we can rewrite the previous equation:
\begin{equation}
    G^{(2l)}(0;2n) = \frac{\mathrm{i}}{b_{2n}}\sum_{\lambda \vdash 2l} \frac{(2l)!}{\prod_{m=1}^{2l}(m!)^{\lambda_{m}}(\lambda_{m})!}(-1)^{|\lambda|}|\lambda|!\frac{\prod_{m=1}^{2l}\Bigl(\partial^{m}_z(p^2_{2n}(z)+p^2_{2n-1}(z))|_{z=0}\Bigl)^{\lambda_m}}{(p^2_{2n}(0))^{|\lambda|+1}} \hspace{5pt}.
    \label{eqn: Derivative}
\end{equation}
Where the symbol $\lambda \vdash 2l$ indicates that $\lambda$ is a (ordered) partition of the set with $2l$ elements, $\lambda_m$ indicates the size of the $m$-th part of a specific partition $\lambda$, and $|\lambda| = \sum_{i=1}^{2l}\lambda_i$. 

The task is to extract the asymptotic behaviour of $G^{(2l)}(0;2n)$ as $n \to \infty$, keeping $l$ an $\mathcal{O}(1)$ constant. In order to do so, we first need to find the asymptotic series for a generic  derivative of the polynomial in the large $n$ limit. This follows from Eq.~\eqref{eqn: Derivative}, which involves a product over the orders of the derivatives of the polynomials. This will be the focus of the next subsection.

\subsection{Asymptotics of the derivatives of the polynomials}

The asymptotic series of the derivative of the polynomial can be found by extending the well-known Christoffel-Darboux formula \cite{Ismail_2005}
\begin{equation}
b_{q+1}\Bigl[p'_{q+1}(x)p_{q}(x)-p_{q+1}(x)p'_{q}(x)\Bigl] = \sum_{j=0}^{q}p^2_{j}(x)    \hspace{5pt},
\label{eqn: Christoffel-Darboux}
\end{equation}
to include a generic derivative of the polynomial:
\begin{equation}
b_{q+1}\Bigl[p_{q+1}^{(m+1)}(x)p^{(m)}_{q}(x)-p_{q+1}^{(m)}(x)p^{(m+1)}_{q}(x)\Bigl] = \sum_{j=0}^{q}\Biggl[(m+1)\Bigl(p^{(m)}_{j}(x)\Bigl)^2-m p^{(m+1)}_{j}(x)p^{(m-1)}_{j}(x)\Biggl]    \hspace{5pt}.
\label{eqn: Generalised CD formula}
\end{equation}
This formula can be derived by considering two copies of the \( m \)-th derivative of the three-term recurrence relation:
\[
b_{j+1} p^{(m)}_{j+1}(x) = x p^{(m)}_j(x) + m p_j^{(m-1)}(x) - b_j p^{(m)}_{j-1}(x),
\]
\[
b_{j+1} p^{(m)}_{j+1}(y) = x p^{(m)}_j(y) + m p_j^{(m-1)}(y) - b_j p^{(m)}_{j-1}(y).
\]
Next, multiply the first equation by \( p^{(m)}_j(y) \) and the second by \( p^{(m)}_j(x) \), then subtract the two equations. Taking the limit \( y \to x \), and finally summing both sides of the resulting equation yields the formula in Eq.~\eqref{eqn: Generalised CD formula}.

By analysing Eq.~\eqref{eqn: Generalised CD formula}, we see that if we know the large-$q$ behaviour of $p_{q}^{(r)}(x)$ for all $r$ up to and including $m$, then, in principle, we can extract the asymptotic form of $p_{q}^{(m+1)}(x)$. However, it is non-trivial since the $(m+1)$-th derivative of the polynomial appears both on the left-hand side but also inside the summation on the right-hand side of Eq.~\eqref{eqn: Generalised CD formula} (except for the special case $m=0$). We are interested in analysing the specific case $x = 0$ of Eq.~\eqref{eqn: Generalised CD formula}. In this case, the polynomial $p^{(m)}_{q}(0)$ must have its upper index $m$ and lower index $q$ of the same parity; otherwise, it vanishes. This condition slightly simplifies Eq.~\eqref{eqn: Generalised CD formula}.
To illustrate this simplification, consider setting $x = 0$, $q=2n$ and $m=2l$ in Eq.~\eqref{eqn: Generalised CD formula}: 
\begin{eqnarray}
 b_{2n+1}p_{2n+1}^{(2l+1)}(0)p^{(2l)}_{2n}(0) &=& \sum_{j=0}^{n}\Biggl[(2l+1)\Bigl(p^{(2l)}_{2j}(0)\Bigr)^2 - 2l\, p^{(2l+1)}_{2j-1}(0) p^{(2l-1)}_{2j-1}(0)\Biggl] \hspace{5pt}.
 \label{eqn: Generalisation}
\end{eqnarray}
Similar expressions can be obtained for all possible parity combinations of $q$ and $m$ in Eq.~\eqref{eqn: Generalised CD formula}.

The task now is to determine the leading order term in the asymptotic series of the derivative of the polynomials by asymptotically balancing both sides of Eq.~\eqref{eqn: Generalisation}.
For ease of notation, let $p^{(2l)}_{2n}(0)$, and similarly $p^{(2l+1)}_{2n+1}(0)$, have the asymptotic series  
$p^{(2l)}_{2n}(0) = \sum_{u \geq 0} p^{(2l)}_{2n,u}$,  
where $u$ labels the order of the asymptotic expansion in $n$.
The first task is to find $p^{(2l)}_{2n,0}$ and $p^{(2l+1)}_{2n+1,0}$. In order to do so, we need to specify the leading order asymptotic of $b_n$. In the following discussion, we focus on the case where the Lanczos coefficients obey the OGH in $d>1$, meaning that $b_n \sim n$.
In this case, it can be shown that the following ansatz
\begin{equation}
    p_{2n,0}^{(2l)} = \frac{(-1)^{n-l}}{2^{2l}\sqrt{2\pi \rho(0)}} \frac{(\log n)^{2l}}{\sqrt{n}} \hspace{30pt} p_{2n+1,0}^{(2l+1)} = \frac{(-1)^{n-l}}{2^{2l+1}\sqrt{2\pi \rho(0)}} \frac{(\log n)^{2l+1}}{\sqrt{n}}
    \label{eqn: Leading order asymptotic m-derivative}
\end{equation}
balances Eq.~\eqref{eqn: Generalisation} to leading order, where $\rho(0)$ is the spectral function at the origin.
This ansatz is motivated by the emergence of a pattern in the zeroth and first derivative.
Consider the following asymptotic 
\begin{equation}
    p_{2n}(0) = (-1)^n\prod_{j=1}^{n}\frac{b_{2j-1}}{b_{2j}} \sim \frac{(-1)^n}{\sqrt{\pi \rho(0) b_{2n}}} \hspace{5pt},
\end{equation}
which follows from the fact that the product in Eq.~\eqref{eqn: Appendix Product} converges by Assumption~\ref{Convergence assumption}. Using Eq.~\eqref{eqn: Generalisation} with $l=0$, we can extract $p^{(1)}_{2n+1,0}$ by simply inverting the formula:
\begin{equation}
p_{2n+1,0}^{(1)} \sim \frac{(-1)^{n}}{2\sqrt{2\pi \rho(0)}} \frac{\log n}{\sqrt{n}}  \hspace{5pt}.  
\end{equation}
The appearance of the logarithm motivates the ansatz presented in Eq.~\eqref{eqn: Leading order asymptotic m-derivative}. Furthermore, the phase is fixed by considering the fact that the coefficients of the polynomials have definite sign, which follows by induction from the three-term recurrence relation.

When we substitute this leading order asymptotic in the expression for $G^{(2l)}(0,2n)$ (Eq.~\eqref{eqn: Derivative}) we find that each term in the summation scales as $(\log n)^{2l}$ (see Subsection~\ref{Cancellation of the divergent term}). This is diverging; however, after summing all the terms, this divergence actually cancels out as we show in Subsection~\ref{Cancellation of the divergent term}. This means that, in order to determine the leading order behaviour of  Eq.~\eqref{eqn: Derivative}, we need the subleading corrections. At the moment, we have an expression for the leading order terms $p^{(2l)}_{2n,0}$ and $p^{(2l+1)}_{2n+1,0}$ which are given by Eq.~\eqref{eqn: Leading order asymptotic m-derivative}; the goal is to find the correction terms $p^{(2l)}_{2n,1}$ and $p^{(2l+1)}_{2n+1,1}$.
Consider the expression in Eq.~\eqref{eqn: Generalisation}, we can obtain an equation for the subleading term by perturbing the left-hand side and the summands on the right-hand side:

\begin{equation}
    b_{2n+1} \Biggl[p^{(2l+1)}_{2n+1,0}p^{2l}_{2n,1} + p^{(2l+1)}_{2n+1,1}p^{(2l)}_{2n,0}  \Biggl] \sim \sum_{j}^{n} \Bigl[2(2l+1)p^{(2l)}_{2j,0}p^{(2l)}_{2j,1} -2l(p^{(2l+1)}_{2j-1,0}p^{(2l-1)}_{2j-1,1}+p^{(2l+1)}_{2j-1,1}p^{(2l-1)}_{2j-1,0}) \Bigl] \hspace{5pt}.
    \label{eqn: Correction}
\end{equation}
We can then asymptotically match the $n$-dependence on both sides to extract the subleading term. 
In general, to determine  $p^{(2l)}_{2n,1}$ and $p^{(2l+1)}_{2n+1,1}$ we need to specify the subleading behaviour of $b_n$.
For the sake of simplicity, we are going to consider staggered correction to the Lanczos coefficient of logarithmic type $s_n = (\log n)^{-a}$. In order for the spectral function to be finite, we must have $a>1$  (see Eq.~\eqref{eqn: Product Asymptotic}) \footnote{In principle, we could generalise the argument to include more generic forms of the staggering; for example, $s_n = (\log n)^{-a} (\log \log\ n)^{-b}$}. This specific choice of staggering is motivated by the fact that for the Freud weight with spectral function $\rho(x) \propto e^{-\pi|x|}$, the Lanczos coefficients have asymptotic form $b_n = n/2 + (-1)^n/(2\log n)^{2} + o((\log n)^{-2})$ \cite{10.1155/S1073792899000161}. In this case, the Green's function has the following small $z$ expansion $G(z)= \alpha + \beta z \log z $ which has a divergent derivative ($k=1$).

As before, we seek an ansatz that consistently balances both sides of Eq.~\eqref{eqn: Generalisation}.  
To find such an ansatz, we calculate the corrections to the zeroth derivative, $p^{(0)}_{2n,1}$, and the first derivative, $p^{(1)}_{2n+1,1}$. By analysing these corrections, we aim to identify a pattern that will guide the formulation of a more general ansatz.
Using the fact that $p_{2n}(0)$ is related to the product in Eq.~\eqref{eqn: Appendix Product}, and applying Eq.~\eqref{eqn: Product Asymptotic}, we obtain the following simple formula:  
\begin{equation}  
    p_{2n,1} \propto \frac{(-1)^{n}}{\sqrt{2\pi \rho(0)}} \frac{1}{\sqrt{n}}\int \frac{\mathrm{d}n}{n}s_n  \hspace{5pt}.
    \label{eqn: Correction of polynomial}  
\end{equation}  
As a side note, the integral should be interpreted ignoring any additive constants.

We can then use Eq.~\eqref{eqn: Correction} with $l = 0$ to estimate the correction to the derivative by simply inverting the formula. This works because the correction to the first derivative, $p^{(1)}_{2n+1,1}$, only appears outside the summation (i.e it only appears on the left-hand side of Eq.~\eqref{eqn: Correction}). We obtain the following expression:
\begin{equation}
    p^{(1)}_{2n+1,1} \propto (-1)^{n}\frac{1}{\sqrt{n}}\left[\int_{\mathcal{O}(1)}^{n}\frac{dn_1}{n_1}\int_{\mathcal{O}(1)}^{n_1}\frac{dn_2}{n_2} s_{n_2} + c_1\right]
    \label{eqn: Correction of 1st derivative of the polynomial} \hspace{5pt}.
\end{equation}
The double integral in this expression should be understood ignoring the constant of integration arising from the first integral over the variable $n_1$. In this expression, we have explicitly included the additive constant $c_1$, which is generally present. It is important to keep track of this constant because, depending on how fast $s_n$ decays, $c_1$ may dominate the first term in brackets.
When staggering is of logarithmic type $s_n = (\log n)^{-a}$ with $a>1$, these integrals can be analytically computed, leading to the simpler forms:
\begin{equation}
p^{(0)}_{2n,1} = A_0(-1)^{n}\frac{(\log n)s_n}{\sqrt{n}} =  A_0(-1)^{n}\frac{(\log n)^{1-a}}{\sqrt{n}}\hspace{30pt} p^{(1)}_{2n+1,1} = A_1(-1)^{n}\frac{(\log n)^{2}s_n+c_1}{\sqrt{n}} = A_1(-1)^{n}\frac{(\log n)^{2-a}+c_1}{\sqrt{n}} \hspace{5pt},
\label{eqn: Pattern}
\end{equation}
where we have introduced some irrelevant constants $A_0$ and $A_1$. Note that since $s_n = (\log n)^{-a}$ with $a>1$ (so that the spectral function is finite at the origin), $p_{2n,1}$ and $p_{2n+1,1}$ are indeed subleading compared to $p_{2n,0}$ and $p_{2n+1,0}$ (see Eq.~\eqref{eqn: Leading order asymptotic m-derivative}). We can observe an emerging pattern in Eq.~\eqref{eqn: Pattern}; therefore, we propose the following ansatz
\begin{equation}
    p^{(2l)}_{2n,1} = (-1)^{n-l}A_{2l}\frac{(\log n)^{2l+1-a}}{\sqrt{n}}
    \hspace{20pt} p^{(2l+1)}_{2n+1,1} = (-1)^{n-l}A_{2l+1}\frac{(\log n)^{2l+2-a}}{\sqrt{n}}
    \hspace{5pt}.
    \label{eqn: Correction derivative}
\end{equation}
This ansatz balances both sides of Eq.~\eqref{eqn: Correction} when the staggered term is of the form $s_n = (\log n)^{-a}$ and $a$ not integer. The simple reason why this ansatz works is that the summation in the right-hand side of Eq.~\eqref{eqn: Correction} can be turned into an integral, we then have to compute integrals of the form:
\begin{equation}
\int  \frac{\mathrm{d}n}{n} (\log n)^{-m} = \frac{(\log n)^{-m+1}}{-m+1} + c  \hspace{40pt} m \neq 1 \hspace{5pt}. 
\end{equation}
Essentially the integration multiplies the integrand by a constant times $n \log n$, which in turns precisely balance what is on the left-hand-side of Eq.~\eqref{eqn: Correction}. This reasoning is correct provided that we do not encounter the special case $m=1$:
\begin{equation}
    \int \frac{\mathrm{d}n}{n} (\log n)^{-1} = \log \log n +c
\end{equation}
for which the answer is not simply asymptotically equivalent to multiplying the integrand by $n \log n$. 
This special case occurs when the right hand side of Eq.~\eqref{eqn: Correction} is $\mathcal{O}(1)$. This can happen after substituting in Eq.~\eqref{eqn: Correction} the asymptotic forms in Eq.~\eqref{eqn: Correction derivative} and Eq.~\eqref{eqn: Leading order asymptotic m-derivative}. In particular, this is the case when the value of $a$ in $s_n = (\log n)^{-a}$ is an even integer satisfying $2l+1 = a/2$. If this is the case, the ansatz in Eq.~\eqref{eqn: Pattern} does not generally balance both sides of Eq.~\eqref{eqn: Correction}. 
However, the expressions for $p^{(0)}_{2n,1}$, Eq.~\eqref{eqn: Correction of 1st derivative of the polynomial}, and $p^{(1)}_{2n+1,1}$, Eq.~\eqref{eqn: Correction of polynomial}, are still valid regardless of the value of $a$. This follows because $p^{(0)}_{2n,1}$ can be determined directly by analysing the infinite product of $b_n$'s (see Eq.~\eqref{eqn: Product Asymptotic}). On the other hand,  $p^{(1)}_{2n+1,1}$ is determined directly from Eq.~\eqref{eqn: Correction} (setting $l=0$) by a simple inversion, since it only appears on the left-hand side.
Notice that a simple way to avoid the edge case $2l+1 = a/2$ is to restrict $a$ from being an integer. If $a$ is not an integer, the ansatz in Eq.~\eqref{eqn: Pattern} always asymptotically balances both sides of Eq.~\eqref{eqn: Correction}. In the next subsection, we focus on the case where $a$ is not an integer, ensuring that the ansatz in Eq.~\eqref{eqn: Pattern} remains valid. We then argue how to extend the criterion to integer values of $a$.
\subsection{Criterion}
\label{Criterion}
In this subsection, we focus on deriving a criterion for the decay of the staggering term, $s_n \propto (\log n)^{-a}$, in dimensions $d > 1$, building on the results from earlier subsections. This criterion provides a range of possible values for $a$ in $s_n = (\log n)^{-a}$ such that the $k$-th derivative of $G(z)$ at the origin is the first to diverge.  
Finally, we state the corresponding criterion for the case $d = 1$, which can be obtained by adapting the calculation for $d > 1$.
\subsubsection{\texorpdfstring{Criterion in $d>1$}{}}
Having obtained expressions for the first two terms in the asymptotic series of the derivative of the polynomials, we can now determine the asymptotic behaviour of $G^{(2l)}(0,2n)$ in Eq.~\eqref{eqn: Derivative}. We remind the reader that $G^{(2l)}(0,2n)$ is expressed as a summation over partitions as presented in Eq.~\eqref{eqn: Derivative}, which we compactly re-write in the following form
\begin{equation}
G^{(2l)}(0;2n) = \sum_{\lambda \vdash 2l} M_{\lambda}L_\lambda (n)
\label{eqn: Sum over partitions}
\end{equation}
where we have defined
\begin{equation}
M_\lambda =  \mathrm{i} \frac{(2l)!}{\prod_{m=1}^{2l}(m!)^{\lambda_{m}}(\lambda_{m})!}(-1)^{|\lambda|}|\lambda|!
\hspace{5pt},
\end{equation}
and 
\begin{equation}
    L_{\lambda}(n) = \frac{\prod_{m=1}^{2l}\Bigl(\partial^{m}_z(p^2_{2n}(z)+p^2_{2n-1}(z))|_{z=0}\Bigl)^{\lambda_m}}{b_{2n}(p^2_{2n}(0))^{|\lambda|+1}}
    \label{eqn: Product of derivative} \hspace{5pt}.
\end{equation}
Recall that $\lambda$ indicates a partition of $2l$ elements, so it is implicitly understood that $\lambda $ depends on $2l$ (i.e. $\lambda = \lambda(2l)$). Recall also that $\lambda_m$ indicates the size of the $m$-th part of a specific partition $\lambda$, and $|\lambda| = \sum_{i=1}^{2l}\lambda_i$. 
It is useful to introduce an additional variable for the factor in Eq.~\eqref{eqn: Product of derivative}:
\begin{equation}
    Q_{m}(n) = \partial^{m}_z\Bigl(p^2_{2n}(z)+p^2_{2n-1}(z)\Bigl)\Bigl|_{z=0} = \sum_{j=0}^{m}\binom{m}{j}\Bigl(p^{(m-j)}_{2n}(0)p^{(j)}_{2n}(0)+p^{(m-j)}_{2n-1}(0)p^{(j)}_{2n-1}(0)\Bigl)
    \label{eqn: Factor} \hspace{5pt}.
\end{equation}
Notice that when $m$ is odd the expression for $Q_m(n)$ is exactly zero. This follows because an odd-order derivative of an even function is an odd function, which vanishes at the origin. Hence, when $m$ is odd we are forced to pick $\lambda_m = 0$, in order to have a non-zero contribution to $L_{\lambda}(n)$ in Eq.~\eqref{eqn: Product of derivative}.
Consequently, let's consider $m$ to be even. Expanding the expression for $Q_{m}(n)$ results in a summation of different terms, as shown in the right-hand side of Eq.~\eqref{eqn: Factor}. We recall that, as we will show in Subsection~\ref{Cancellation of the divergent term}, the right-hand of Eq.~\eqref{eqn: Factor} vanishes if we substitute all instances of the derivatives of the polynomials with their leading-order terms:
\begin{equation}
\sum_{j=0}^{m}\binom{m}{j}\Bigl(p^{(m-j)}_{2n,0}p^{(j)}_{2n,0}+p^{(m-j)}_{2n-1,0}p^{(j)}_{2n-1,0}\Bigl) = 0   \hspace{5pt}.
\end{equation}
Thus, we must introduce the subleading terms in the asymptotic series of the derivatives of the polynomials (see Eq.~\eqref{eqn: Correction derivative}). Therefore, the large $n$-behaviour of $Q_m(n)$ is given by:
\begin{equation}
    Q_m(n) \sim 2 \sum_{j=0}^{m}\binom{m}{j}\Bigl(p^{(m-j)}_{2n,0}p^{(j)}_{2n,1} +p^{(m-j)}_{2n-1,0}p^{(j)}_{2n-1,1} \Bigl) \propto
    \frac{(\log n)^{m+1-a}}{n} \hspace{5pt}.
    \label{eqn: Factor Asymptotic}
\end{equation}
To get this asymptotic behaviour of $Q_m(n)$, we assume that the prefactor does not vanish anomalously.
Plugging this into Eq.~\eqref{eqn: Product of derivative}, yields the following asymptotics
\begin{equation}
    L_{\lambda}(n) = \frac{\prod_{m=1}^{2l}(Q_m(n))^{\lambda_m}}{b_{2n}(p^2_{2n}(0))^{|\lambda|+1}} \propto (\log n)^{2l+(1-a)|\lambda|} \hspace{5pt}.
    \label{eqn: L_n}
\end{equation}
Recall that $a>1$; hence, the leading order term is achieved for the minimum value of $|\lambda|$, which corresponds to a partition $\lambda^{*}$ such that $\lambda^{*}_{2l} = 1$ and $\lambda^{*}_j = 0$ when $j \neq 2l$. Hence, we conclude that 
\begin{equation}
    G^{(2l)}(0;2n) \propto L_{\lambda^{*}}(n) \propto (\log n)^{2l+1-a} + C_{2l} \hspace{5pt}.
\end{equation}
We have explicitly written $C_{2l}$ to indicate an additive constant that is bounded in $n$; this constant arises because of the additive constant present in Eq.~\eqref{eqn: Correction of 1st derivative of the polynomial}.  It is important to keep track of this constant because, depending on the values of $l$ and $a$, it may dominate the first term, leading to a competition between the two contributions.  
This constant is generally present, since by  assumption we have that:
\begin{equation}
    \lim_{n \to \infty}G^{(k)}(0;2n) = G^{(k)}(0) \hspace{5pt}.
\end{equation}
Hence, by assumption, when $|G^{(2l)}(0)|<\infty$ the constant $C_{2l}$ dominates the other term $(\log n)^{2l+1-a}$ and the limit is finite.
For the sake of simplicity, we have restricted the analysis to the even derivative. However, it can be shown that a similar result holds also for odd derivatives. Hence, we conclude that the generic asymptotic behaviour of \( G^{(k)}(0;n) \) is given by:  
\begin{equation}
    G^{(k)}(0;n) \propto (\log n)^{k+1-a} + C_{k} \hspace{5pt}.
    \label{eqn: Asymptotics of G^k}
\end{equation}  
This equation is significant as it reveals a competing mechanism. In particular, the asymptotic behaviour depends on the sign of \( k+1-a \):  
\begin{itemize}  
    \item If \( k+1-a > 0 \), the first term dominates \( C_k \), leading to \( G^{(k)}(0;n) \) diverging.  
    \item If \( k+1-a < 0 \), the first term vanishes asymptotically, and \( G^{(k)}(0;n) \) converges to \( C_k \).  
\end{itemize}  

Requiring \( G^{(k)}(0;n) \) to diverge while keeping \( G^{(k-1)}(0;n) \) finite imposes a condition on the decay of the staggered term \( s_n \) in the Lanczos coefficients. If the \( k \)-th derivative of the Green’s function diverges while the \((k-1)\)-th derivative remains finite, then \( s_n \) must satisfy  
\begin{equation}  
(\log n)^{k+1} s_n = (\log n)^{k+1-a}  \to \infty, \hspace{20pt} (\log n)^k s_n =  (\log n)^{k-a} < \infty.  
\label{eqn: Appendix Criterion}  
\end{equation}  
This implies the condition \( k < a < k+1 \), where \( a \) is the decay rate of the staggered term, given by \( s_n = (\log n)^{-a} \).  

Recall that, in the previous subsection, we derived expressions for the correction to the derivative of the polynomials, given by Eq.~\eqref{eqn: Correction derivative}. However, these expressions are generally valid only when \( a \) is not an integer. In the case where \( a \) is an integer, we only have valid expressions for the zeroth and first derivatives.
Consequently, it is not a priori clear whether the extrema $a = k$ and $a = k+1$ correspond to a divergent $k$-th derivative, as our analysis is restricted to integer values of $a$.
To address the edge case when $a$ is integer, we propose the following extension of the previous criterion. If $G^{(k)}(0)$ is the first divergent derivative, the subleading staggered term in the Lanczos coefficient $s_n$ should satisfy:
\begin{eqnarray}   
   &\phantom{=}&  (\log n)^{k-1} \int_{\mathcal{O}(1)}^{n}\frac{\mathrm{d}n_1}{n_1}\int_{\mathcal{O}(1)}^{n_1}\frac{\mathrm{d}n_2}{n_2} s_{n_2} \to \infty   \hspace{30pt} (\log n)^{k-2} \int_{\mathcal{O}(1)}^{n}\frac{\mathrm{d}n_1}{n_1}\int_{\mathcal{O}(1)}^{n_1}\frac{\mathrm{d}n_2}{n_2}s_{n_2}  < \infty
    \label{eqn: Appendix k-1-differentiability} \hspace{5pt}.
\end{eqnarray}
This version of the criterion agrees with Eq.~\eqref{eqn: Appendix Criterion} whenever $a$ is not integer. Furthermore, it agrees with the well-known case of the Freud weight with spectral function $\rho(x) \propto e^{-\pi|x|}$. In this case, the Lanczos coefficients have asymptotic form $b_n = n/2 + (-1)^n/(2\log n)^{2}$ \cite{10.1155/S1073792899000161}, and the first derivative is the first divergent derivative (i.e. $k=1$). Setting $k=1$ and $s_n = 1/(2\log n)^{2}$ in Eq. \eqref{eqn: Appendix k-1-differentiability}, yields the first term diverging as $\log \log n$, whereas the second term does not diverge.

To motivate this criterion when $a$ is integer, we will make the assumption that $G^{(2l)}(0;2n)$ asymptotically is captured by those terms in the right-hand side of Eq.~\eqref{eqn: Sum over partitions} that involves the correction of the first derivative, $p^{(1)}_{2n+1,1}$. This is motivated by the fact that the expression for $p^{(1)}_{2n+1,1}$ in Eq.~\eqref{eqn: Correction of 1st derivative of the polynomial} is valid even when $a$ is integer. Keeping track of these terms implies that $G^{(2l)}(0;2n)$ contains a term that asymptotically scales as
\begin{equation}
    (\log n)^{2l-1}\int_{\mathcal{O}(1)}^{n}\frac{\mathrm{d}n_1}{n_1}\int_{\mathcal{O}(1)}^{n_1}\frac{\mathrm{d}n_2}{n_2} s_{n_2}
    \label{eqn: Term j=1} \hspace{5pt},
\end{equation}
which is precisely the first term in Eq.~\eqref{eqn: Appendix k-1-differentiability} when $k=2l$.  

We now explain how this term arises in more details by considering $Q_m(n)$, given in Eq.~\eqref{eqn: Factor}. Recall that $Q_m(n)$ is a factor appearing in the expression for $G^{(2l)}(0;2n)$ (see Eq.~\eqref{eqn: Sum over partitions} and Eq.~\eqref{eqn: Factor}). We can see that the term $p^{(1)}_{2n-1,1}p^{(m-1)}_{2n-1,0}$ is present as a summand ($j=1$) in Eq.~\eqref{eqn: Factor Asymptotic}, this term scales as:
\begin{equation}
    p^{(1)}_{2n-1,1}p^{(m-1)}_{2n-1,0} \propto \frac{(\log n)^{m-1}}{n} \int_{\mathcal{O}(1)}^{n}\frac{\mathrm{d}n_1}{n_1}\int_{\mathcal{O}(1)}^{n_1}\frac{\mathrm{d}n_2}{n_2} s_{n_2} \hspace{5pt}.
\end{equation}
When this term is substituted into Eq.~\eqref{eqn: L_n}, using the specific partition $\lambda^{*}$, it exactly reproduces the term in Eq.~\eqref{eqn: Term j=1}. Therefore, this analysis shows that the term in Eq.~\eqref{eqn: Term j=1} is explicitly present in the expression for $G^{(2l)}(0;2n)$. 

\subsubsection{\texorpdfstring{Criterion in $d=1$}{}}
So far we have focused the discussion on the case of the OGH in $d>1$. It can be shown that when $b_n = n/\log n$ ($d=1$), the analysis is very similar expect that the asymptotic for the $m$-th derivative of the polynomial is slightly different:
\begin{equation}
    p_{2n,0}^{(2l)} = \frac{(-1)^{n-l}}{2^{4l}\sqrt{2\pi \rho(0)}} \frac{(\log n)^{4l+1/2}}{\sqrt{n}} \hspace{30pt} p_{2n+1,0}^{(2l+1)} = \frac{(-1)^{n-l}}{2^{2(2l+1)}\sqrt{2\pi \rho(0)}} \frac{(\log n)^{2(2l+1)+1/2}}{\sqrt{n}}
    \hspace{5pt}.
\end{equation}
As a result, the ansatz for the correction is
\begin{equation}
p^{(2l)}_{2n,1} =  A_{2l} \frac{(-1)^{n-l}}{2^{4l} \sqrt{2\pi \rho(0)}} \frac{(\log n)^{4l+5/2-a}}{\sqrt{n}}  \hspace{30pt} p^{(2l+1)}_{2n+1,1} = A_{2l+1} \frac{(-1)^{n-l}}{2^{2(2l+1)} \sqrt{2\pi \rho(0)}} \frac{(\log n)^{2(2l+1)+5/2-a}}{\sqrt{n}}
\hspace{5pt}.
\end{equation}
Similarly to the previous case, we have assumed that the subleading staggered correction to the Lanczos coefficient is of logarithmic type $s_n \propto (\log n)^{-a}$. In this case, we require that $a>2$, in order for the spectral function to be finite at the origin (see Eq.~\eqref{eqn: Criterion}).
These expressions lead to the following criterion for the $k$-th derivative to diverge and the previous ones to be finite in $d=1$
\begin{equation}
(\log n)^{2k+2}s_n \to \infty \hspace{20pt} (\log n)^{2k}s_n < \infty
\label{eqn: Criterion d=1} \hspace{5pt};
\end{equation}
this condition is met when $2k<a<2(k+1)$.

We then propose the following integral form of the criterion in $d=1$, which covers the case when $a$ is integer:
\begin{align}   
   \phantom{=}&  (\log n)^{2(k-1)} \int_{\mathcal{O}(1)}^{n}\mathrm{d}n_1\frac{\log n_1}{n_1}\int_{\mathcal{O}(1)}^{n_1}\mathrm{d}n_2\frac{\log n_2}{n_2} s_{n_2} \to \infty  \nonumber \\
   \phantom{=}& (\log n)^{2(k-2)} \int_{\mathcal{O}(1)}^{n}\mathrm{d}n_1\frac{\log n_1}{n_1}\int_{\mathcal{O}(1)}^{n_1}\mathrm{d}n_2\frac{\log n_2}{n_2}s_{n_2}  < \infty
    \label{eqn: d=1, k-1-differentiability} \hspace{5pt}.
\end{align}
\subsection{Cancellation of the divergent term}
\label{Cancellation of the divergent term}
In this part, we provide additional details on why it is necessary to compute the corrections to the leading-order terms of the derivatives of the polynomials ($p^{(2l)}_{2n,1}$ and $p^{(2l+1)}_{2n+1,1}$). As previously mentioned, the reason is that a cancellation occurs in the expression for $G^{(k)}(0;n)$ when only the leading-order terms, $p^{(2l)}_{2n,0}$ and $p^{(2l+1)}_{2n+1,0}$, are substituted. 

To show this explicitly, we analyse the asymptotic form, as $n\to \infty$, of Eq.~\eqref{eqn: Product of derivative}:
\begin{equation}
\frac{1}{b_{2n}}\frac{\prod_{m=1}^{2l}\Bigl(\partial^{m}_z(p^2_{2n}(z)+p^2_{2n-1}(z))|_{z=0}\Bigl)^{\lambda_m}}{(p^2_{2n}(0))^{|\lambda|+1}} = \frac{\prod_{m=1}^{2l}\Bigl[\sum_{j=0}^{m}\binom{m}{j}\Bigl(p^{(m-j)}_{2n}(0)p^{(j)}_{2n}(0)+p^{(m-j)}_{2n-1}(0)p^{(j)}_{2n-1}(0)\Bigl)\Bigl]^{\lambda_m}}{b_{2n}(p_{2n}(0))^{2(|\lambda|+1)}}  \hspace{5pt}.
\label{eqn: Product_Cancellation}
\end{equation}
Recall that $\lambda_m$ is a nonnegative integer, and that we consider nonnegative integers satisfying $\sum_{m=1}^{2l}m \lambda_m =2l$, we also use the notation $|\lambda | = \sum_{i=1}^{2l}\lambda_i$. 
Focus first on the two terms inside the brackets:
\[p^{(m-j)}_{2n}(0)p^{(j)}_{2n}(0)+p^{(m-j)}_{2n-1}(0)p^{(j)}_{2n-1}(0) \hspace{5pt}.\]
In order to have nonzero contributions to the product, the following restrictions on $j$ should be imposed depending on the parity of $m$:
\begin{itemize}
    \item $m$ even: $j$ should be even in the first term and odd in the second term 
    \item $m$ odd: in this case no matter the parity of $j$ both terms always vanish. Hence, we must have $\lambda_m = 0$ so that the product in Eq.~\eqref{eqn: Product_Cancellation} does not automatically vanish.
\end{itemize}
Hence, we parametrise $j \to 2j_1$ (even) in the first term and $j \to 2j_2+1$ (odd) in the second term. 
Notice that the first term, $p^{(m-2j_1)}_{2n}(0)p^{(2j_1)}_{2n}(0)$,
and the second term, $p^{(m-2j_2-1)}_{2n-1}(0)p^{(2j_2+1)}_{2n-1}(0)$, have the same leading order asymptotics (up to signs and prefactors) regardless of what $j_1$, $j_2$ actually are. This follows from the ansatz in Eq.~\eqref{eqn: Leading order asymptotic m-derivative}, which we restate below (for $d>1)$:
\begin{equation}
    p_{2n,0}^{(2v)} = \frac{(-1)^{n-v}}{2^{2v}\sqrt{2\pi \rho(0)}} \frac{(\log n)^{2v}}{\sqrt{n}} \hspace{30pt} p_{2n+1,0}^{(2v+1)} = \frac{(-1)^{n-v}}{2^{2v+1}\sqrt{2\pi \rho(0)}} \frac{(\log n)^{2v+1}}{\sqrt{n}} \hspace{5pt}.
    \label{eqn: Derivatives leading order reminder}
\end{equation}
It follows that $p^{(m-2j_1)}_{2n}(0)p^{(2j_1)}_{2n}(0)$ and $p^{(m-2j_2+1)}_{2n-1}(0)p^{(2j_2-1)}_{2n-1}(0)$ are both $\mathcal{O}\left((\log n)^{m}/n\right)$, when $m$ is even. Therefore, asymptotically, a generic term in Eq.~\eqref{eqn: Product_Cancellation} scales as
\begin{equation}
\frac{1}{b_{2n}}\frac{\prod_{m=1}^{2l}\Bigl[\frac{(\log n)^{m}}{n}\Bigl]^{\lambda_m}}{(p_{2n}(0))^{2(|\lambda|+1)}} = \mathcal{O}\left( \frac{1}{n}\frac{(\log n)^{2l}}{n^{|\lambda|}}\left(\sqrt{n}\right)^{^{2(|\lambda|+1)}}\right) =  \mathcal{O}((\log n)^{2l}) \hspace{5pt},
\end{equation}
where we have used the facts that $\sum_{m=1}^{2k}m \lambda_m =2l$ and $|\lambda | = \sum_{i=1}^{2l}\lambda_i$.

We conclude that a generic term in the expression for $G^{(2l)}(0;2n)$ in Eq.~\eqref{eqn: Derivative} is asymptotically proportional to $(\log n)^{2l}$; this is divergent. However, we will show that this divergence actually cancels out.
To do this, we need to track signs to show that the cancellation occurs. Luckily, we do not need to perform the sum over all partitions in Eq.~\eqref{eqn: Derivative}, the cancellation occurs in this part of the expression (see Eq.~\eqref{eqn: Product_Cancellation}):
\[
\prod_{m=1}^{2l}\Bigl[\sum_{j=0}^{m}\binom{m}{j}\Bigl(p^{(m-j)}_{2n}(0)p^{(j)}_{2n}(0)+p^{(m-j)}_{2n-1}(0)p^{(j)}_{2n-1}(0)\Bigl)\Bigl]^{\lambda_m} \hspace{5pt}.
\]
Recall that if $m$ is odd $\lambda_m = 0$, if $m$ is even then $j$ must be even in the first term and odd in the second term inside the brackets. We will now show that all the factors corresponding to even $m$ vanish, and therefore the entire product vanishes.
In this case, we get
\begin{align*}
    & \sum_{j_1=0}^{m/2}\binom{m}{2j_1}\Bigl(p^{(m-2j_1)}_{2n}(0)p^{(2j_1)}_{2n}(0)\Bigl)+\sum_{j_2=0}^{m/2-1}\binom{m}{2j_2+1}\Bigl(p^{(m-2j_2-1)}_{2n-1}(0)p^{(2j_2+1)}_{2n-1}(0)\Bigl)   \\
    \sim & \sum_{j_1=0}^{m/2}\binom{m}{2j_1}\Bigl(\frac{(-1)^{\frac{m}{2}}}{2^{m}\cdot 2\pi \rho(0)} \frac{(\log n)^{m}}{n}\Bigl) + \sum_{j_2=0}^{m/2-1}\binom{m}{2j_2+1}\Bigl(\frac{(-1)^{\frac{m}{2}+1}}{2^{m}\cdot 2\pi \rho(0)} \frac{(\log n)^{m}}{n}\Bigl) \\
    = & \frac{(-1)^{\frac{m}{2}}}{2^{m}\cdot 2\pi \rho(0)} \frac{(\log n)^{m}}{n}\left(\sum_{j_1=0}^{m/2}\binom{m}{2j_1}-\sum_{j_2=0}^{m/2-1}\binom{m}{2j_2+1}\right) = \frac{(-1)^{\frac{m}{2}}}{2^{m}\cdot 2\pi \rho(0)} \frac{(\log n)^{m}}{n}\left(\sum_{j=0}^{m}\binom{m}{j}(-1)^{j}\right) = 0 \hspace{5pt}.
\end{align*}
In the first line, we have separated the summands based on the parity of $j$, in the second line we have used the leading order term in the asymptotic of the derivative of the polynomials (see Eq.~\eqref{eqn: Leading order asymptotic m-derivative}).

This calculation shows that the factor is zero when $m$ is even, and consequently, if we substitute only the leading-order asymptotic terms in Eq.~\eqref{eqn: Product_Cancellation}, a cancellation occurs.

\subsection{Numerics}
We now present numerical evidence to support the asymptotic results for $G^{(k)}(0; n)$ discussed in Section~\ref{Criterion}.  
In general, we find that higher-order derivatives are more challenging to fit accurately; hence, we restrict our attention to the first two derivatives.  
We consider four different sets of Lanczos coefficients: two for the case $d = 1$ and two for $d > 1$, since the asymptotic form of $G^{(k)}(0; n)$ differs between these regimes.
\subsubsection{\texorpdfstring{$d=1$}{d=1}}

For $d=1$, the asymptotic behaviour of $G^{(k)}(0; n)$ is predicted to follow the form
\begin{equation}
    G^{(k)}(0;n) \propto (\log n)^{2(k-1)} \int_{\mathcal{O}(1)}^{n} \frac{\log n_1}{n_1} \, \mathrm{d}n_1 \int_{\mathcal{O}(1)}^{n_1} \frac{\log n_2}{n_2} \, \mathrm{d}n_2 \, s_{n_2} + C_k \,,
    \label{Prediction G d=1}
\end{equation}
where $s_n$ is the subleading staggered correction in the Lanczos coefficients, and $C_k$ is a constant that may dominate the first term depending on the value of $k$ and the decay rate of $s_n$.

We test this prediction using the following Lanczos coefficients, which follow the OGH in $d=1$:
\begin{enumerate}
    \item $b_n = \frac{n}{\log(n+1)} + 1 + \frac{1}{10} (-1)^n (\log(n+1))^{-3}$. 
    \item $b_n = \frac{n}{\log(n+1)} + \frac{(-1)^n}{12} (\log(n+1))^{-5}$.
\end{enumerate}

From these coefficients, we numerically calculate the first and second derivatives: $G^{(1)}(0;n)$ and $G^{(2)}(0;n)$. The scaling behavior predicted by Eq.~\eqref{Prediction G d=1} is summarized in Table~\ref{tab: Prediction d=1}.

\begin{table}[H]
    \centering
    \footnotesize
    \scalebox{1}{
    \begin{tabular}{c@{\hspace{2.5em}}c@{\hspace{2.5em}}c}
        \toprule \\ [-0.7em]
        Lanczos coefficients ($d=1$) & $G^{(1)}(0;n)$ & $G^{(2)}(0;n)$ \\
        [0.2em] \hline & \\[1em] 
        \textbf{Case 1:} $b_n = \frac{n}{\log(n+1)} + 1 + \frac{(-1)^n}{10} (\log(n+1))^{-3}$  & $\log n$ & $(\log n)^3$\\
        [2em]
        \textbf{Case 2:} $b_n = \frac{n}{\log(n+1)} + \frac{(-1)^n}{12}(\log(n+1))^{-5}$ & $\mathcal{O}(1)$ & $\log n$\\ [1em] \\
        \botrule
    \end{tabular}}
    \caption{Asymptotic behaviour of $G^{(1)}(0;n)$ and $G^{(2)}(0;n)$ for the Lanczos coefficients $b_n$ listed in the leftmost column, which obey the OGH in $d=1$.}
    \label{tab: Prediction d=1}
\end{table}

As shown in Table~\ref{tab: Prediction d=1}, Case 1 yields a divergent first derivative, since $G^{(1)}(0;n) \propto \log n$ diverges as $n \to \infty$. In Case 2, the first derivative remains bounded, $G^{(1)}(0;n) = \mathcal{O}(1)$, while the second derivative diverges as $G^{(2)}(0;n) \propto \log n$.

To numerically confirm these asymptotic forms, we compute $G^{(1)}(0;n)$ and $G^{(2)}(0;n)$ for sufficiently large values of $n$. This is necessary for two reasons. First, $n$ must be large enough to enter the asymptotic regime. Second, $n$ must be large enough to clearly distinguish logarithmic growth from a plateau. The predictions in Table~\ref{tab: Prediction d=1} are supported by the numerical evaluations shown in Subfigures (a) and (b) of Fig.~\ref{fig: Cases G first derivative} and Fig.~\ref{fig: Cases G second derivative}.

Therefore, we conclude that the spectral function corresponding to Case 1 has a divergent derivative at the origin. In contrast, the spectral function for Case 2 has a finite first derivative but a divergent second derivative.
\subsubsection{\texorpdfstring{$d>1$}{d>1}}
\begin{figure}
        \centering
        \includegraphics[scale=0.1405]{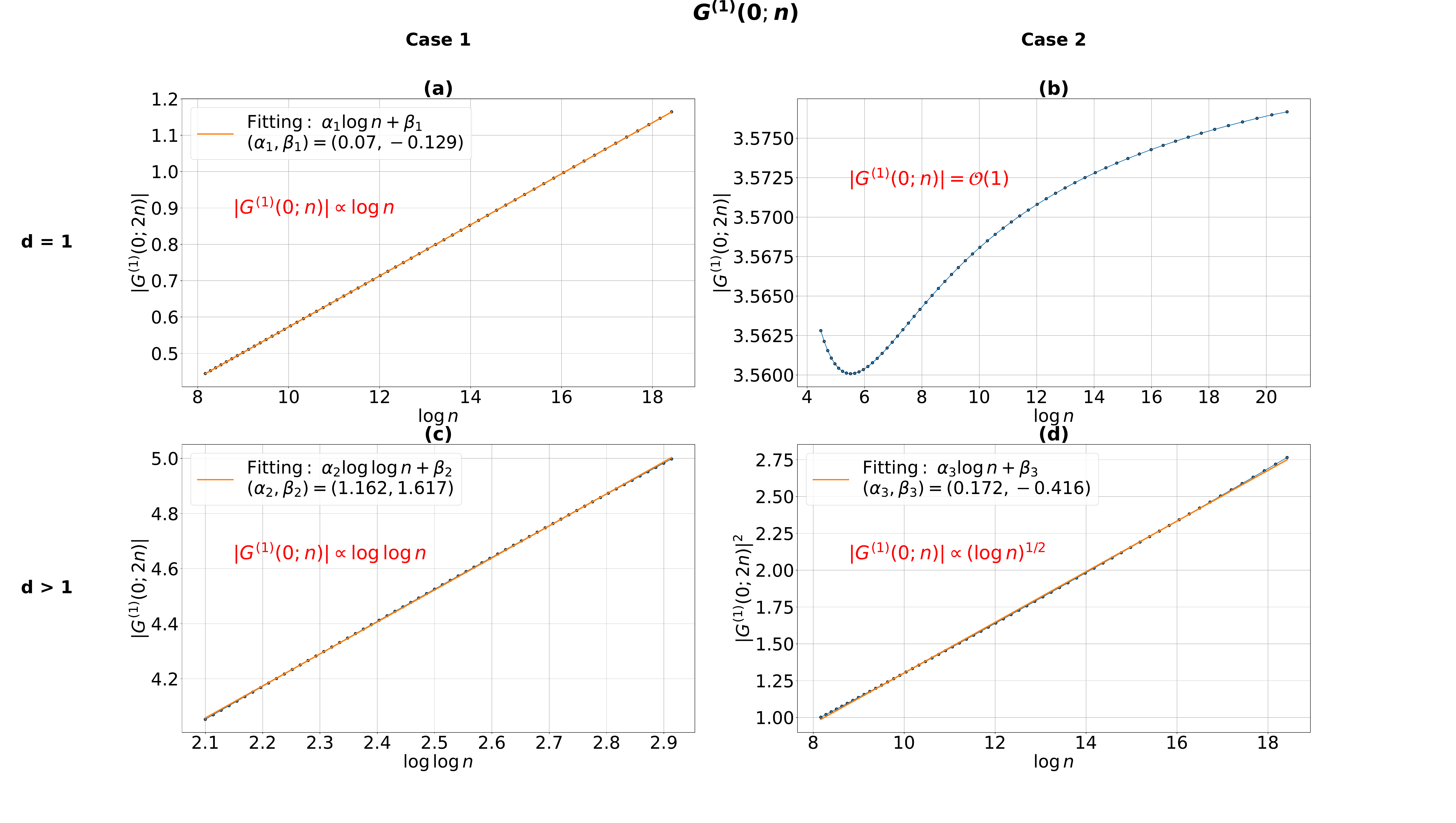}
    \caption{Numerical evaluations of $G^{(1)}(0; 2n)$ for the different Lanczos coefficients listed in Table~\ref{tab: Prediction d=1} ($d=1$) and Table~\ref{tab: Prediction d>1} ($d>1$). The red annotations in each plot indicate the theoretical prediction for the scaling of $G^{(1)}(0; n)$. In all cases except Subfigure~(b), $G^{(1)}(0;n)$ diverges as $n \to \infty$. To test the theoretical prediction in these cases, we present suitably rescaled quantities on the horizontal and vertical axes such that, if the prediction holds, the data should align along a straight line. For each of these cases, we also include the best fitting line for comparison.}
    \label{fig: Cases G first derivative}
\end{figure}

\begin{figure}
        \centering
        \includegraphics[scale=0.1405]{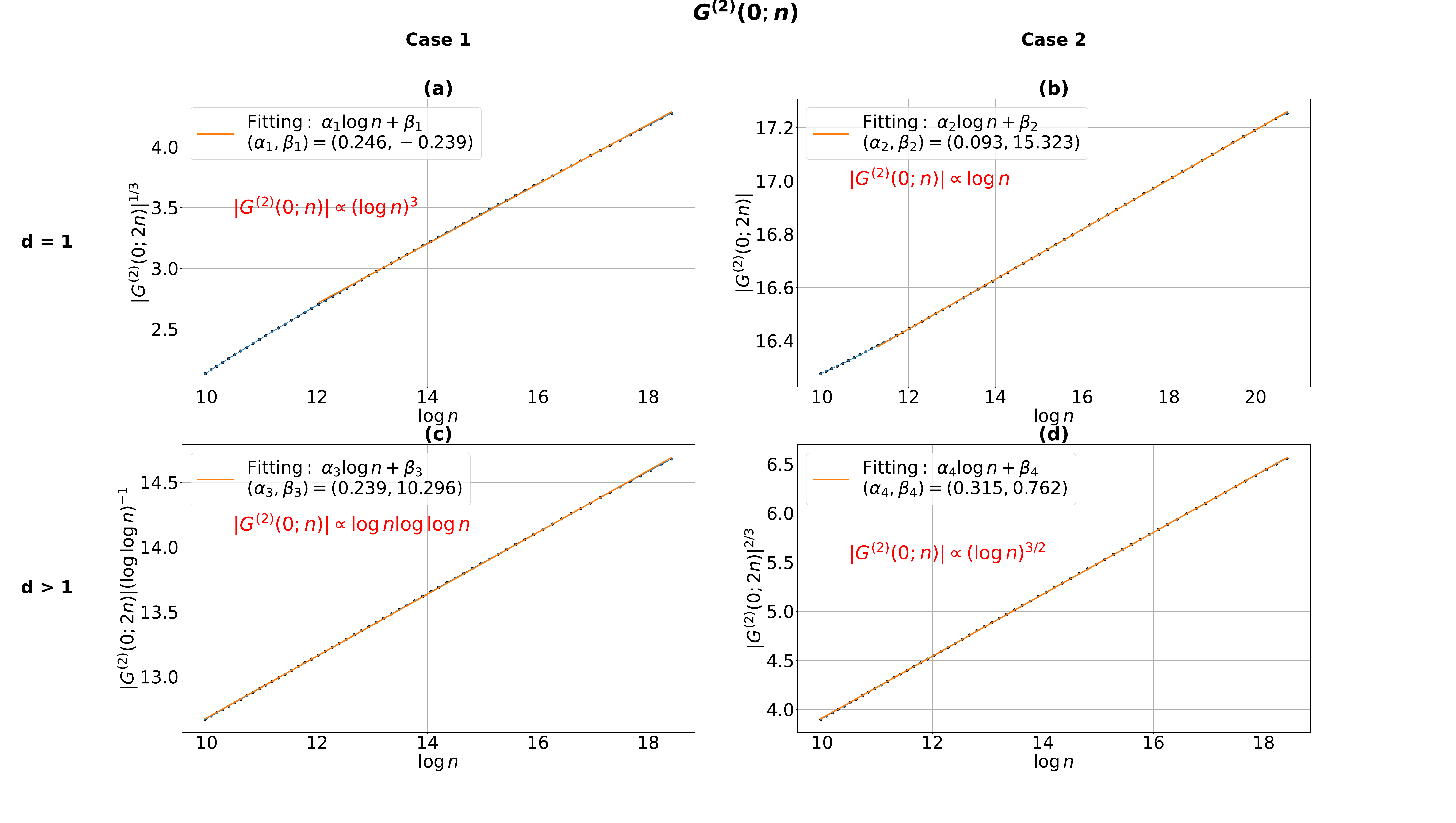}
    \caption{Numerical computations of $G^{(2)}(0; 2n)$ for the different Lanczos coefficients listed in Table~\ref{tab: Prediction d=1} ($d=1$) and Table~\ref{tab: Prediction d>1} ($d>1$). The red labels in each plot indicate the theoretical prediction for the scaling of $G^{(2)}(0; n)$. In all cases $G^{(2)}(0;n)$ is predicted to diverge as $n \to \infty$. To test the rate of divergence, we display rescaled quantities on the horizontal and vertical axes such that, if the prediction holds, the data should align along a straight line. In each case, a best-fit line is also shown for comparison.}
    \label{fig: Cases G second derivative}
\end{figure}

We now turn to the case $d > 1$, where the asymptotic behaviour of $G^{(k)}(0;n)$ takes on a slightly different form:
\begin{equation}
    G^{(k)}(0;n) \propto (\log n)^{k-1}\int_{\mathcal{O}(1)}^{n}\frac{\mathrm{d}n_1}{n_1}\int_{\mathcal{O}(1)}^{n_1}\frac{\mathrm{d}n_2}{n_2} s_{n_2} + C_k \hspace{5pt}.
     \label{Prediction G d>1}
\end{equation}
To evaluate this prediction, we consider the following Lanczos coefficients, which obey the OGH in $d>1$:
\begin{enumerate}
    \item $b_n = n + 1 + \frac{(-1)^n}{4} (\log(n+1))^{-2}$ 
    \item $b_n = n + 1 + \frac{(-1)^n}{10} \left( \log(n+1) \right)^{-3/2}$.
\end{enumerate}
The first case involves an integer power of the staggering term $(\log n)^{-2}$, which, as noted in Section~\ref{Criterion}, represents an edge case. The second case, by contrast, involves a non-integer power, $(\log n)^{-3/2}$, and is considered a regular case.
The prediction from 
Eq.~\eqref{Prediction G d>1} is presented in Table~\ref{tab: Prediction d>1}.
\begin{table}[H]
    \centering
    \footnotesize
    \scalebox{1}{
    \begin{tabular}{c@{\hspace{2.5em}}c@{\hspace{2.5em}}c}
        \toprule \\ [-0.7em]
        Lanczos coefficients ($d>1$) & $G^{(1)}(0;n)$ & $G^{(2)}(0;n)$ \\
        [0.2em] \hline & \\[1em] 
        \textbf{Case $1$:} $b_n = n + 1 + \frac{(-1)^n}{4} (\log(n+1))^{-2}$  & $\log \log n$ & $\log n\log \log n$\\
        [2em]
        \textbf{Case $2$:} $ b_n = n + 1 + \frac{(-1)^n}{10} \left( \log(n+1) \right)^{-3/2}$ & $(\log n)^{1/2}$ & $(\log n)^{3/2}$\\ [1em] \\
        \botrule
    \end{tabular}}
    
    \caption{Asymptotic scaling of $G^{(1)}(0;n)$ and $G^{(2)}(0;n)$ depending on the two cases of the Lanczos coefficients presented in the leftmost column which obey the OGH in $d>1$.}
    \label{tab: Prediction d>1}
\end{table}
The predictions in Table~\ref{tab: Prediction d>1} generally agree with the numerical evaluation of $G^{(1)}(0;n)$ and $G^{(2)}(0;n)$ shown in Subfigures (c) and (d) of Fig.~\ref{fig: Cases G first derivative} and Fig.~\ref{fig: Cases G second derivative}. The only caveat is Case 1 in $d > 1$, presented in Subfigure (c) of Fig.~\ref{fig: Cases G first derivative}. In this case, the data are  not fully conclusive, as the range of $n$ is not wide enough to clearly resolve the expected $\log \log n$ scaling. Nevertheless, the growth is slow and clearly sub-logarithmic, consistent with the theoretical prediction.
\end{document}